\documentclass[a4paper,fleqn,usenatbib]{mnras}

\usepackage[T1]{fontenc}
\usepackage{ae,aecompl}
\usepackage{rotating}

\usepackage{graphicx}	% Including figure files
\usepackage{amsmath}	% Advanced maths commands
\usepackage{amssymb}	% Extra maths symbols
\usepackage{caption}
\usepackage{subcaption}

\title[K2-106]{A new mass and radius determination of the ultra-short period planet K2-106b and the fluffy planet K2-106c \thanks{Based on observations made with ESO Telescopes at the La Silla Paranal Observatory under programme 0103.C-0289(A).}}
\author[Eike W. Guenther et al.]
  {E.W. Guenther$^{1}$\thanks{E-mail: guenther@tls-tautenburg.de},
  E. Goffo$^{1,2}$,
  D. Sebastian $^{3}$,
  A.M.S. Smith$^{4}$,
  C.M. Persson$^{5}$,
     \newauthor 
  M. Fridlund$^{5,6}$,
  D. Gandolfi$^{2}$,
  J. Korth$^{7}$ \\
\\
$^{1}$ Th\"uringer Landessternwarte Tautenburg, Sternwarte 5, 07778
  Tautenburg, Germany. \\
$^{2}$ Dipartimento di Fisica, Universita degli Studi di Torino, via Pietro Giuria 1, I-10125, Torino, Italy.\\
$^{3}$ 
School of Physics and Astronomy University of Birmingham, Birmingham University, Edgbaston Park Rd, Birmingham B15 2TT, UK \\
$^{4}$ Institute of Planetary Research, German Aerospace Center (DLR),  Rutherfordstrasse 2, 12489 Berlin, Germany \\
$^{5}$ Department of Space, Earth and Environment, Chalmers University of Technology, Onsala Space Observatory, SE-439 92 Onsala, Sweden.\\
$^{6}$ Leiden Observatory, University of Leiden, PO Box 9513, 2300 RA, Leiden, The Netherlands. \\
$^{7}$ Lund Observatory, Division of Astrophysics, Department of Physics, Lund University, Box 118 , 22100 Lund, Sweden.
}

\date{Accepted 2024, February 14. Received  2023 January 20}

\pubyear{2024}

\begin{document}
\label{firstpage}
\pagerange{\pageref{firstpage}--\pageref{lastpage}}
\maketitle


\begin{abstract}  
Ultra-short period planets have orbital periods of less than one day. 
Since their masses and radii can be determined to a higher precision than long-period planets, they are the preferred targets to determine the density of planets which constrains their composition.
The K2-106 system is particularly interesting because it contains two planets
of nearly identical masses. 
One is a high density USP, the other is a low-density planet that has an orbital period of 13 days. 
Combining the Gaia DR3 results with new ESPRESSO data allows us to determine the masses and radii of the two planets more precisely than before. 
We find that the USP K2-106\,b has a density consistent with an Earth-like composition, and K2-106\,c is a low-density planet that presumably has an extended atmosphere.
We measure a radius of  $\rm R_p=1.676_{-0.037}^{+0.037}$ $\rm R_{\oplus}$, a mass of  $\rm M_p=7.80_{-0.70}^{+0.71}$  $ M_{\oplus}$ and a density of $\rm \rho=9.09_{-0.98}^{+0.98}$  
$\rm g\,cm^{-3}$ for K2-106\,b.
For K2-106\,c, we derive $ R_p=2.84_{-0.08}^{+0.10}$ $\rm R_{\oplus}$, $M_p=7.3_{-2.4}^{+2.5}$ $\rm M_{\oplus}$, and a density of  $\rm \rho= 1.72_{-0.58}^{+0.66}$ $\rm g\,cm^{-3}$. 
We finally discuss the possible structures of the two planets with  respect to other low-mass planets.  
\end{abstract}
  


%%%%%%%%%%%%%%%%%%%%%%%%%%%%%%%%%%%%%%%%%%%%%%%%%%

%%%%%%%%%%%%%%%%% BODY OF PAPER %%%%%%%%%%%%%%%%%%

\section{Introduction}
\label{sectI}

Ultra-short period planets (USPs) are an enigmatic subset of exoplanets with  orbital periods less than one day. USPs are the focus of many research programs because their masses and radii can easily be determined with high precision. This stems from large radial-velocity (RV) amplitudes  and easy detection of a large number of transits. The first USP, and the first rocky exoplanet discovered, was CoRoT-7\,b \citep{2009AuA...506..287L}. 

Low-mass USPs (lmUSPs) are particularly interesting to study, because they can not have extended hydrogen atmospheres. This is because any hydrogen atmosphere would be quickly eroded by X-ray and extreme UV radiation from the host star \citep{2017AuA...598A..90F, 2018A&A...619A.151K}.  They are thus often referred to as bare rocks. Measurements of their densities thus allow us to draw conclusions about the  internal structure of rocky planets. Planets without an H/He atmosphere can have masses up to 25 $\rm M_{\oplus}$ \citep{2020A&A...634A..43O}. We therefore define lmUSPs as planets with $\rm P_{orb} < 1$~day, and $\rm Mp < 25 M_{\oplus}$.  
Currently 35 USPs have been discovered in the mass range between 1 and 25 $\rm M_{\oplus}$. 

Another reason to study lmUSPs is that their formation is still being debated with several possible scenarios  \citep{2021ApJ...919...26U}. 
One scenario is that lmUSPs are the remnant cores of gas-giants that lost their atmospheres due to photo-evaporation, or Roche-lobe overflow \citep{2014RSPTA.37230164M,2020Natur.583...39A}. 
TOI-849\,b  has been suggested to be a remnant of a gas giant, because it has as mass of $39.1_{-2.6}^{+2.7}$ $\rm M_{\oplus}$, and a density of $\rm \rho=5.2_{-0.8}^{+0.7}$  g\,cm$^{-3}$ \citep{2020Natur.583...39A}. 
However, not all gas-giant USPs evaporate, as has been shown by the discoveries of WASP-18b \citep{2009Natur.460.1098H}, WASP-19b \citep{2010ApJ...708..224H}, and NGTS-10b \citep{2020MNRAS.493..126M}. 

A second scenario is that lmUSPs have developed through mass accretion in the innermost part of the protoplanetary disk \citep{2019AJ....157..180P}.  
Because lmUSPs should be bare rocks, one would expect that they should all have densities consistent with the abundances of rock forming elements of their host stars. 
However that is not the case, it looks like that lmUSPs are more diverse. 
The theoretical mass-radius-relation has recently been derived for rocky and for volatile rich planets by \citet{2020A&A...634A..43O}.

We can define three classes of low-mass planets: 

\begin{itemize}
\item [A)] Planets with densities that are lower than that of a planet with Earth-like core to-mantle-ratio.
\item [B)] Planets whose density is consistent with an Earth-like core-to-mantle ratio.
\item [C)] Planets with densities that are higher than for a planet with an Earth-like core-to-mantle ratio.
\end{itemize}

Given that the composition of planets is only inferred from the density measurement, we prefer to classify the planets based only on their density, rather than their inferred composition.

Gas-giant USPs fall into class-A but WASP-18b, WASP-19b and NGTS-10b have more than $25 M_{\oplus}$, and are thus not lmUSPs. 
Examples for class-A lmUSPs are 55\,Cnc\,e  \citep{2018RNAAS...2..172C}, and WASP-47\,e \citep{2017AJ....154..237V}. 
They can also have masses below two $M_{\oplus}$ \citep{2022arXiv220707456L}.
Class-A lmUSPs can also have longer orbital periods.
Examples for class-A planets that are not USPs are K2-3\,b,c \citep{ 2018A&A...615A..69D,2019AJ....157...97K,2022arXiv220712755D}, and HD\,219134\,b \citep{
2015ApJ...814...12V,2015A&A...584A..72M,2019A&A...631A..92L, 2019MNRAS.484..712D, 2023AuA...677A..12D}. 

There are mainly two possibilities what class-A planet could be.
One is that they contain low density material like water ice, or Aluminium-rich minerals.
\citet{2019MNRAS.484..712D} showed that Ca-, Al-rich minerals may be enhanced in lmUSPs that formed close to the star. 

Another possibility is that class-A planets have hybrid atmospheres \citep{2023arXiv230110217T}. 
The outgassing hypothesis is plausible, because USPs are likely to have lava oceans \citep{2010ASPC..430..409B, 2010ApJ...709L..95B}. 
Close-in rocky planets could also have exospheres like Mercury \citep{2008E&PSL.271..181E, 2011Icar..211....1M}.

There is indirect evidence that at least some class-A planets could have an atmosphere.
Infrared observations of 55\,Cnc\,e show that the hottest point is not at the substellar point, but east of it. 
This can best be explained by an atmosphere \citep{2017AJ....154..232A}.
\citet{2016A&A...593A.129R} and \citet{2016ApJ...820...99T} claimed to have detected the atmosphere directly, but this was not confirmed by \citet{2017AJ....153..268E} and \citet{2020MNRAS.498.4222T}. 
HST observations of $\rm \pi\,Men\,c$ showed that it has a hybrid atmosphere \citep{2021ApJ...907L..36G}.
In contrast to 55\,Cnc\,e, phase curves of the lmUSP K2-141\,b do not show any significant thermal hotspot offset. A rock vapour model and a 1D turbulent boundary layer model both fits well to the observations \citep{2022A&A...664A..79Z}.

At the present stage it is not known what the structure and composition of
class-A planets is.  We thus prefer to define the planets on the basis of their density, rather than composition.

Because class-C planets have a high density, they must have relatively large cores. 
However, the main issue for many UPSs is that the density measurements are not precise enough to conclude whether they are class-B, or class-C.

Possibly the best case for a class-C planet is the USP GJ367\,b \citep{2021Sci...374.1271L, 2023ApJ...955L...3G}.
Although \citet{2022MNRAS.513..661B} obtained a smaller mass and larger radius for the planet than \citet{2021Sci...374.1271L} and \citet{2023ApJ...955L...3G}, this planet still falls in to class-C. 

Other examples for class-C planets are
K2-229\,b, \citep{2018NatAs...2..393S}, and HD80653\,b  \citep{2020AuA...633A.133F}.
Although the masses of KOI 1843.03 and K2-137b have not been determined yet, the Roche limit implies mean densities of $\rm \rho_p> 7\,g\,cm^{-3}$ for KOI 1843.03 
\citep{2013ApJ...773L..15R}, and $\rm \rho_p>6.4\,g\,cm^{-3}$ for K2-137b
\citep{2018MNRAS.474.5523S}, respectively. KOI 1843.03 and K2-137b must therefore be class-C planets. A disputed case is K2-106\,b.

Like class-A planets, class-C planets can also have orbital periods longer than one day.
An example for a class-C planet that is not a USP is HD\,137496\,b \citep{2022A&A...657A..68A}.  

There is thus evidence that class-C planets exist but how did they form?
An interesting case is the Kepler-107 system, because the inner planet is in class-B, and the outer in class-C.
Because of this unusual architecture, \citet{2019NatAs...3..416B} argue that Kepler-107\,c is the result of a giant impact that removed the outer layers of the planet.

Does this mean that all class-C planets are the result of impact stripping? 
This is not the case.
\citet{2022MNRAS.517.3132R} investigated how mantle stripping by giant impacts changes the composition of planets.
\citet{2022EPSC...16...74A} showed that the iron-mass fraction of planets is on average higher than that of the primordial values,
and \citet{2020MNRAS.493.4910S} studied the composition of rocky exoplanets in the context of the composition of the stars.  
The result is that the composition of rocky planets spans a much wider range than that of stars.
Super-Mercuries and super-Earths appear to be two distinct groups of planets.
Mantle stripping alone thus can not explain all class-C planets, formation and stripping both play a role.

Previous studies thus have shown that all three classes exist amongst USPs as well as for planets with longer periods. 
Studying USPs has the advantage that we can measure their masses and radii more precisely.
USPs may have atmospheres but extended H/He atmospheres have not yet been found amongst lmUSPs.

Unfortunately, masses and radii of only a few USPs have been determined accurately enough to categorize them \citep{2020MNRAS.499..932P}.
The most recent radius- and mass-distribution of USPs has been published by \citet{2021ApJ...919...26U}. 
However, not only accurate measurements are important, it is also important which theoretical mass-radius relation is used. 
For example, it makes a difference whether we use the relation
from \citet{2011Icar..214..366W}, \citet{2018Icar..313...61H}, or \citet{2019PNAS..116.9723Z}.
Thus, up to now there are only few planets that can be firmly categorised.
Any additional object is important.

K2-106\,b is a particularly interesting planet, because it orbits a relatively bright star, it is one the most massive rocky USPs known, and contradicting density measurements have been published
\citep{2017AJ....153..271S,2017AuA...608A..93G,2019ApJ...883...79D, 2022AuA...658A.132S, 2023AJ....165...97R}. It could thus either be of class-A, or B. 
K2-106\,b was discovered by \citet{2017AJ....153...82A} using K2-data. 
Previous mass, radius and density values that were derived for  K2-106\,b are given in Table\,\ref{tab:planetI}.

\begin{table*}
\caption{Radius, mass, and density of K2-106\,b from the literature}
\begin{tabular}{l l l l}
\hline
$\rm Rp$    & $\rm Mp$    & density        & reference \\
$\rm [R_{\oplus}]$ & $\rm [M_{\oplus}]$ & $\rm [g\,cm^{-3}]$ &  \\
\hline
$\rm {1.82}^{+0.20}_{-0.14}$ & $\rm 9.0\pm1.6$ & $\rm {8.57}^{+4.64}_{-2.80}$ & \cite{2017AJ....153..271S} \\
$\rm 1.46\pm0.14$   & - & - & \cite{2017AJ....153...82A} \\
$\rm {1.52}\pm0.16$ & $\rm {8.36}^{+0.96}_{-0.94}$ & $\rm {13.1}^{+5.4}_{-3.6}$ & \citet{2017AuA...608A..93G} \\
$\rm 2.31\pm0.16$ & - & - &  \citet{2018AJ....156..277L}\\
$\rm 1.712\pm0.068$ & $\rm {7.72}^{+0.80}_{-0.79}$ & $\rm 8.5\pm1.9$ & \cite{2019ApJ...883...79D} \\
$\rm 1.6\pm0.1$ & - & - &  \citet{2021PSJ.....2..152A} \\
% & & & \citet{2022AuA...658A.132S} \\
% & & & \citet{2022arXiv220807883R} \\
$\rm {1.71}^{+0.069}_{-0.057}$ & $\rm 8.53\pm1.02$ & $\rm {9.4}^{+1.6}_{-1.5}$ & \citet{2023AJ....165...97R} \\
$\rm 1.725\pm0.039$ & $\rm 8.21^{+0.76}_{-0.74}$ & $\rm {8.77}^{+1.00}_{-0.94}$ &  \citet{2023AuA...677A..33B} \\
\hline
\end{tabular}
\label{tab:planetI}
\end{table*}

Another interesting property of K2-106\,b is that it has a very high maximum geometric albedo of $0.9\pm0.3$, and a maximum dayside temperature of $3620^{+56}_{-53}$\,K \citep{2022AuA...658A.132S}. 
A lava ocean of that temperature is expected to have a high albedo \citep{2011ApJ...741L..30R}. 
K2-106\,c has the same mass as K2-106\,b, but a lower density.

The mass and radius measurements can now be significantly improved.
The radius of the star was originally determined using Gaia DR1, or by the analysis of stellar spectra. 
The improvement when using Gaia DR3 compared to DR1 is quite significant.
The Gaia DR1 parallax was $\rm 3.96\pm0.78$\,mas and the parallax from Gaia DR3 is $\rm 4.085\pm0.018$\,mas
\citep{2021A&A...649A...1G}.
Additional transits were observed by TESS, and we have obtained additional spectra with ESPRESSO. 
The ESPRESSO spectra have a higher resolution, a higher S/N, and a much higher radial velocity (RV) accuracy.
The higher resolution and the higher S/N allows to determine the mass, and radius of the star
to a higher accuracy. 
The new data allows to find out, whether K2-106\,b is in class-A, or in class-B. 

\section{Observations and results}
\label{sectIII}

In the following sections we present new determinations of the fundamental parameters of the host star and compare them with previous estimates.

\subsection{Mass, radius and other stellar parameters derived from stellar magnitudes and Gaia DR3 parallax}
\label{sectIIIb}

The combination of Gaia parallax with broad-band photometry allows one to determine the radius of a star \citep[see, e.g.,][]{2018AJ....156..102S}. The parallax of K2-106 is reported to be $\pi\,=\,4.085\pm0.018$\,$\mu$as in the Gaia Data Release 3 \citep[Gaia DR3; ][]{2021A&A...649A...1G}.

\begin{figure}
\includegraphics[height=0.45\textwidth,angle=270.0]{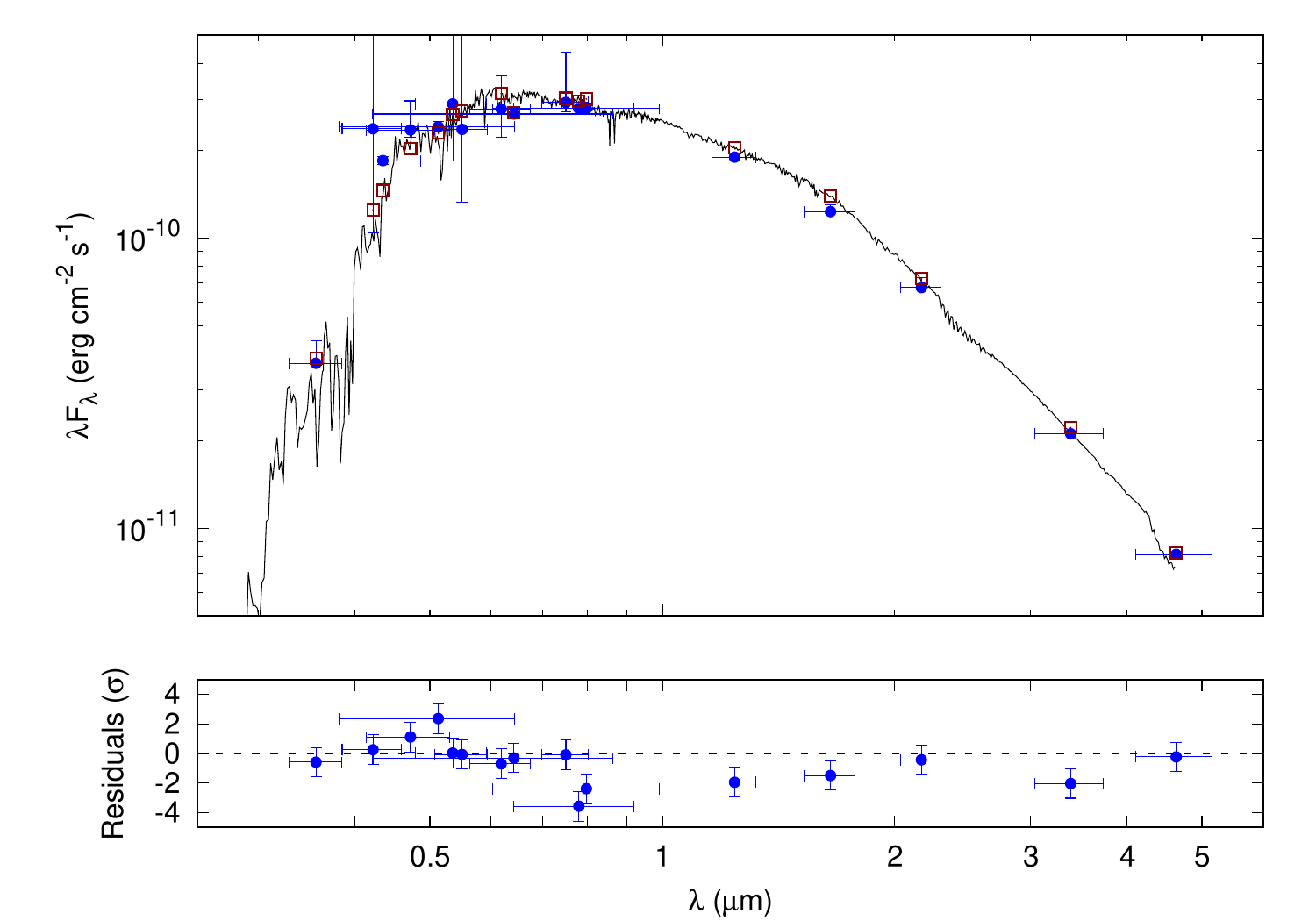}
\caption{Spectral energy distribution and the fit obtained with ARIADNE. The lower panel shows the residuals.}
\label{ARIADNE}
\end{figure}

Using the SED-fitting algorithm ARIADNE \footnote{Available at \url{https://github.com/jvines/astroARIADNE}} \citep[spectrAl eneRgy dIstribution bAyesian moDel averagiNg fittEr;][]{2022MNRAS.513.2719V}, we obtain the parameters given in Table\,\ref{tab:starI}. With ARIADNE we derive $\rm R_{*}=1.039^{+0.025}_{-0.023}$\,R$_\odot$ and
$\rm M_{*}=1.046^{+0.046}_{-0.053}$ using the MIST isochrones.
ARIADNE uses the following magnitudes for K2-106: 2MASS J, H, K, Johnson V, B, Tycho B, V, Gaia G, Rp, Bp, SDSS g', r', i', WISE, W1, W2, and TESS.  Since the SDSS z' magnitude is a clear outlier, we did not include it in the fit.
ARIADNE uses a number of different stellar models
(Phoenix V2 \citep{2005ESASP.576..565B}, BT-Models (\url{https://osubdd.ens-lyon.fr/phoenix/}), and the 
Kurucz models \citep{1993yCat..32810817C,1993yCat.6039....0K}. Shown in Fig. \ref{ARIADNE} are the Kurucz models together with the photometric measurements. The SED fit shows that is a main-sequence star. The values obtained with  are listed in Table\,\ref{tab:starII}. 

\begin{figure}
\includegraphics[height=0.35\textwidth,angle=0.0]{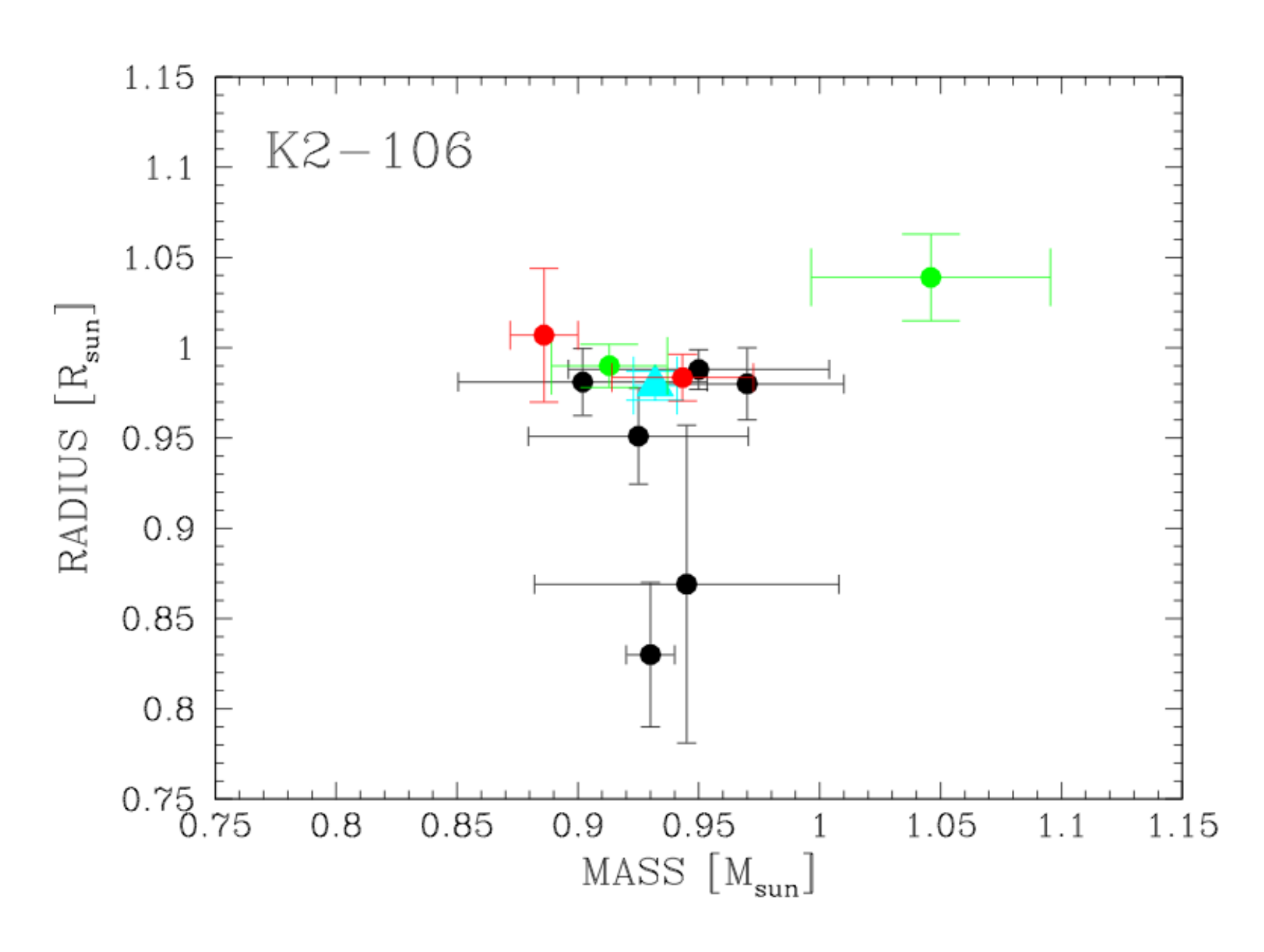}
\caption{Mass and radius of the star. Values from the literature (black), average 
value from the literature (blue triangle), the new value derived from ESPRESSO spectra (red), and the new value from Gaia EDR3 using ARIADNE (MIST isochrones) and ISOCHRONES (green).}
\label{mass_radius_star}
\end{figure}

We also determine the mass and radius, as well as other stellar parameters using the ISOCHRONES code \citep{2015ascl.soft03010M} using with the
MESA isochrones \citep{Dotter16,Choi16}. This method utilities a multimodal nested sampler {\tt multinest} \citep{Feroz08,Feroz09,Feroz19} to sample 1000 live points with following input and errors of the values: $\rm T_{eff}=5493\,K$ [5433 to 5553 K], [Fe/H]=-0.11 [-0.11 to +0.11], parallax: 4.085 mas [4.035 to 4.135 mas] and using the brightness values listed in  Table\,\ref{tab:starIII}. Instead of B and V we used the Gaia values because they are more precise. We also tried out using the B and V instead of the Gia values but did not find a significant difference. 
This global fit takes the respective uncertainties into account, and the value derived can exceed the uncertainty of a specific input value. Figure\,\ref{mass_radius_star} shows the mass and radius-values of the host star obtained in this work with the literature. With ISOCHRONES we derive $\rm R_{*}=0.990\pm0.012$\,R$_\odot$ and $\rm M_{*}=0.913\pm0.024$.

The radius of the star can also be obtained by combining the Gaia parallax and the 2MASS photometry \citep{2006AJ....131.1163S} 
as described in \citet{2021MNRAS.507.2103G}. Using this method, we obtain $\rm R_{*}=0.989\pm0.022$\,R$_\odot$. This method has the advantage that it is less affected
by interstellar absorption, because it uses only infrared colours. However, the extinction should not play a significant role given that the star has a Galactic longitude $l=123.3^o$, the Galactic latitude $b=-52.1^o$, and a distance of $244.8\pm1.1$\,pc.

We also determined the stellar mass and radius using the on-line version of EXOFASTv1 \citep{2013PASP..125...83E}
\footnote{https://exoplanetarchive.ipac.caltech.edu/cgi-bin/ExoFAST/nph-exofast}.
The mass of the star as obtained with $\mathcal{X}^2$ fitting algorithm 
is $\rm M_{*}=0.96\pm0.01\,M{\odot}$ and with the MCMC algorithm we obtained
$\rm M_{*}=0.96_{-0.04}^{+0.05}\,M{\odot}$. For the stellar radius, we derive $\rm R_{*}=0.92\pm0.17\,M_{\odot}$ with $\mathcal{X}^2$ fitting algorithm, and $\rm R_{*}=1.00_{-0.04}^{+0.04}$ with the MCMC.
As shown in Table\,\ref{tab:starI}, these values are also in agreement with other values for mass and radius, derived in this work.

\begin{table*}
\caption{Properties of the host star derived from Gaia data and using from Gaia using 
ARIADNE and from ISOCHRONES}
\begin{tabular}{l c l l}
\hline
\noalign{\smallskip}
Parameter & Gaia & ARIADNE & ISOCHRONES \\   
          &      &  & \\ 
\hline
RA (J2000.0)$^{(1)}$   & 00:52:19.1  & & \\
DE (J2000.0)$^{(1)}$   & +10:47:40.91 & & \\
pm RA [mas/yr]$^{(1)}$ & $61.01\pm0.02$ & &\\
pm RA [mas/yr]$^{(1)}$ & $2.06\pm0.01$ & & \\
RV $\rm [km\,s^{-1}]^{(1)}$ & $-14.79\pm0.56$ & & \\
Parallax [mas]$^{(1)}$ & $4.09\pm0.02 $  & & \\
Distance [pc]	   & $244.8\pm1.1$  &      &        \\
$\rm R_{*}$	[R$_\odot$]   & & $1.039^{+0.025}_{-0.023}$ & $0.990\pm0.012$ \\
$\rm M_{*}$ [M$_\odot$] & & $1.046^{+0.046}_{-0.053}$ $^{(2)}$ & $0.913\pm0.024$      \\
$\rm T_{eff}$ [K]  & & $5493^{+57}_{-61}$ & $5578\pm46$ \\
Luminosity [$\rm L_\odot$]	& & $0.885^{+0.057}_{-0.054}$ &  \\
log(g)	       & & $4.421^{+0.076}_{-0.068}$  $\rm 4.26^{+0.08}_{-0.08}$ & $4.408\pm0.017$ \\
$\rm [Fe/H]$   & & $-0.11^{+0.11}_{-0.11}$  & $-0.005\pm0.059$ \\
Extinction Av  & & $0.098^{+0.040}_{-0.044}$ &  $0.219\pm0.038$             \\
% AD[mas]        & $0.0377^{+0.0012}_{-0.0013}$  &         \\
 \hline
\end{tabular}
\label{tab:starII}
\\
$^{(1)}$ 	Gaia DR3 2582617711154563968; 
$^{(2)}$ Mass derived by interpolating MIST isochrones;
% $^{(3)}$ Noise-term as explained in \citep{2022MNRAS.513.2719V};
\end{table*}

\begin{table}
\caption{Brightness of the star}
\begin{tabular}{l l}
\hline
\noalign{\smallskip}
band  & mag  \\   
\hline
V & $\rm 12.58\pm0.26$ \\
B & $\rm 12.10\pm0.21$ \\
Gaia\,G\  & $\rm 11.9438\pm0.0028$ \\
Gaia\,BP  & $\rm 12.3315\pm0.0028$\\
Gaia\,RP  & $\rm 11.3935\pm0.0038$ \\
Tycho\,B\_T' & $\rm 12.67\pm0.26$ \\
Tycho\,V\_T' & $\rm 12.16\pm0.21$ \\
SDSS\,g'  & $\rm 12.629\pm0.010$ \\
SDSS\,r'  & $\rm 12.0262\pm0.0095$ \\
SDSS\,i'  & $\rm 11.812\pm0.0100$ \\
TESS      & $\rm 11.457\pm0.078$ \\
J         & $\rm 10.77\pm0.023$ \\
H         & $\rm 10.454\pm0.026$ \\
Ks        & $\rm 10.344\pm0.021$ \\
WISE 1    & $\rm 10.299\pm0.023$ \\
WISE 2    & $\rm 10.355\pm0.021$ \\
WISE 3    & $\rm 10.380\pm0.091$ \\
 \hline
\end{tabular}
\label{tab:starIII}
\end{table}

\begin{table*}
\caption{Radius, mass, temperature and log(g) of the host star from the literature and derived in this work}
\begin{tabular}{l l l l r l}
\hline
\noalign{\smallskip}
$\rm R_{*}$ & $\rm M_{*}$ & $\rm T_{\rm eff}$ & $\rm log(g)$ & [Fe/H] & reference \\
$\rm [R\odot]$ & $\rm [M\odot]$ & [K] & & dex &  \\
\hline
% $\rm 0.95\pm0.05$             & $\rm 0.92\pm0.003$  & $\rm 5496\pm46$ & $4.42\pm0.05$ & $\rm -0.047\pm0.002$ & \cite{2017AJ....153..271S} \\
$\rm 0.83\pm0.04$ & $\rm 0.93\pm0.01$ & $\rm 5590\pm51$ & $\rm 4.56\pm0.09$ & $\rm 0.025\pm0.02$ & \cite{2017AJ....153...82A} \\
$\rm 0.869\pm0.088$ & $\rm 0.945\pm0.063$ & $\rm 5470\pm30$ & $\rm 4.53\pm0.08$ &  $\rm -0.025\pm0.05$ & \citet{2017AuA...608A..93G} \\
$\rm 0.98\pm0.02$ & $\rm 0.97\pm0.04$ & $\rm 5617\pm86$ & $\rm 4.45\pm0.03$ &  $\rm 0.13\pm0.06$ & \citet{2018AJ....156..277L}\\
$\rm 0.981^{+0.019}_{-0.018}$ & $\rm 0.902^{+0.057}_{-0.046}$ & $\rm 5496\pm46$ &  $\rm 4.42\pm0.05$ & $\rm 0.06\pm0.03$ & \cite{2019ApJ...883...79D} \\
$\rm 0.95\pm0.05$ & - & $\rm 5613\pm39$ & $\rm 4.60\pm0.07 $ & $\rm 0.01\pm0.01$ & \cite{2021PSJ.....2..152A} \\
$\rm 0.951^{+0.027}_{-0.026}$ & $\rm 0.925^{+0.049}_{-0.042}$ & $\rm 5598^{+80}_{-78}$ & $\rm 4.449^{+0.031}_{-0.029}$ & $\rm 0.096^{+0.060}_{-0.058}$ & \citet{2022AuA...658A.132S} \\
% $\rm 0.99^{+0.029}_{-0.027}$ & $\rm 1.02^{+0.022}_{-0.032}$ & $\rm 5489^{+75}_{-76}$ & $\rm 4.446^{+0.007}_{-0.016}$ & $\rm -0.03\pm0.001$ & \citet{2022arXiv220807883R} \\
% $\rm 0.98^{+0.024}_{-0.022}$ & $\rm 0.96^{+0.063}_{-0.062}$ & $\rm 5508\pm70$ & $\rm 4.44^{+0.033}_{-0.037}$ & $\rm -0.03\pm0.001$ & \citet{2022arXiv220807883R}$^{(1)}$ \\
$\rm 0.988\pm0.011$ & $\rm 0.950^{+0.060}_{-0.048}$ & $\rm 5532\pm62$ & - & $\rm 0.11\pm0.05$ &
\citet{2023AuA...677A..33B} \\
\hline
$\rm 0.979\pm0.008$ & $\rm 0.932\pm0.009$ & $\rm 5535\pm18$ & $\rm 4.46\pm0.02$ & $\rm 0.022\pm0.008$ & average of Literature values \\
\hline
\hline
$\rm 1.007_{-0.024}^{+0.050}$ & $\rm 0.886_{-0.014}^{+0.014}$ & $\rm 5344\pm60$ &  $\rm 4.26^{+0.08}_{-0.08}$ & $\rm -0.03\pm0.05$ & ESPRESSO$^{(1)}$ \\
$\rm 0.983\pm0.013$ & $\rm 0.943\pm0.029$ & $\rm 5488\pm60$ & 
$\rm 4.28\pm0.20 $ & $\rm 0.04\pm0.08$ & ESPRESSO$^{(2)}$ \\
$\rm 1.039^{+0.025}_{-0.023}$ & $\rm 1.05^{+0.19}_{-0.17}$ & $\rm 5493^{+57}_{-61}$ & $\rm 4.421^{+0.076}_{-0.068}$ & $\rm -0.11\pm0.11$ & SED + Gaia DR3$^{(3)}$\\  
$\rm 1.039^{+0.025}_{-0.023}$ & $\rm 1.046^{+0.046}_{-0.053}$ & $\rm 5493^{+57}_{-61}$ & $\rm 4.421^{+0.076}_{-0.068}$ & $\rm -0.11\pm0.11$ & SED + Gaia DR3$^{(4)}$\\  
$\rm 0.989\pm0.022$ & & & & & 2MASS + Gaia DR3$^{(5)}$\\
$\rm 0.990\pm0.012$ & $\rm 0.913\pm0.024$ & $\rm 5578\pm46$ & $\rm 4.408\pm0.017$ & $\rm -0.05\pm0.06$ &  ISOCHRONES \\
\hline
$\rm 0.993\pm0.008$ & $\rm 0.907\pm0.011 $ & $\rm 5491\pm28$ & $\rm 4.402\pm0.016$ & $\rm -0.02\pm0.03$ & average this work \\
\hline
\end{tabular}
\\
% $^{(1)}$ MIST isochrones;
$^{(1)}$ Mass and radius calculated from modelling the ESPRESSO spectra using the Kurucz ATLAS12 models and the MESA isochrones. See Section \ref{sectIIIa} for details.
$^{(2)}$ Mass and radius calculated using ESPRESSO spectra, the FeI, FeII lines and the WIDTH radiative transfer code. See Section \ref{sectIIIa} for details.
$^{(3)}$ Mass calculated directly from the star’s log g and $\rm R_{*}$; 
$^{(4)}$ Mass derived by interpolating MIST isochrones; 
$^{(5)}$ Method described in \citet{2021MNRAS.507.2103G}.
\label{tab:starI}
\end{table*}

\subsection{Mass, radius and other stellar parameters derived from ESPRESSO spectra}
\label{sectIIIa}

We acquired 23 spectra of K2-106 using the ESPRESSO spectrograph \citep{2014AN....335....8P, 2021A&A...645A..96P} at the VLT UT3 (Melipal) as part of program 0103.C-0289(A). The spectra were obtained from August $8^{th}$ 2019 to November $16^{th}$ 2019.  We used the high-resolution mode which gives a resolving power of $\rm \lambda/\Delta\lambda \sim 140\,000$. The spectra cover the wavelength range from 3782\,\AA\, to 7887\,\AA.  All calibration frames were taken using the standard procedures of this instrument.  The spectra were reduced and extracted using the dedicated ESPRESSO pipeline.

We derived new values for $\rm T_{eff}$, log(g) and [Fe/H] using the six ESPRESSO spectra with the highest S/N ratio using the same method as described in \citet{2023MNRAS.tmp.3690O,2023MNRAS.526..548O,2023AuA...677A..12D,2023A&A...674A.117G,2023MNRAS.519.1437L,2022A&A...666A.184P,2022NatAs...6..736S,2021MNRAS.505.4684G, 2020MNRAS.498.4503F,2019AcA....69..135S}.
We fixed the micro- and macro-turbulence to $\rm v_{mic} = 0.9\pm0.1$\,km\,s$^{-1}$ and $\rm v_{mac} = 1.70\pm0.35$\,km\,s$^{-1}$, using the relations given by \citet{2014MNRAS.444.3592D} and \citet{2010MNRAS.405.1907B}, respectively. Fitting the $\rm H_{\alpha}$ profile using 
Kurucz ATLAS12 models and spectra models, \citep{2013ascl.soft03024K} we find $\rm T_{eff}=5344\pm60\,$K.
This corresponds to a spectral type G9\,V, according to the  mean dwarf stellar color and effective temperature sequence\footnote{Available at \url{https://www.pas.rochester.edu/~emamajek/EEM_dwarf_UBVIJHK_colors_Teff.txt}.} from \citet{2012ApJ...746..154P}. 

The projected rotation velocity of the star is $\rm v\,sin\,i_\star\,=\,2.7\pm0.4\,km\,s^{-1}$, which gives a statistical age of $\rm 1.3_{-0.3}^{+0.6}$\,Gyrs \citep{2022A&A...663A.142M}. The v\,sin\,i has been determined using unblended FeI lines.
The activity index $\rm log\,(R'_{HK})$ was already published in \citet{2017AuA...608A..93G}.  On average it is $\rm log\,(R'_{HK})=-5.04\pm0.19$, which gives a statistical age of $\rm 7.4_{-3.4}^{+3.0}$\,Gyrs \citep{2008ApJ...687.1264M}. Based on CaIIHK-index K2-106 is an old, inactive star. 

At first glance it appears the that relatively rapid rotation of the star contradicts the low $\rm log\,(R'_{HK})$ index and the old age derived. 
However, the stellar spin-down can be affected by close-in planets
\citep{2019A&A...621A.124B,2022MNRAS.513.4380I,2023RAA....23i5014G}. A star hosting a close-in planet 
thus may rotate faster than a star without close-in planets.

Many planet host stars also have an abnormally low level 
Ca\,II\,H\&K emission which is interpreted as a signature of atmospheric mass-loss from planets 
rather than a low activity level, or an old age of the host star \citep{2012ApJ...760...79H, 2020NatAs...4..399S, 2023MNRAS.524.5196B}. Given that both, the rotation velocity, and the Ca\,II\,H\&K flux can be affected by a close-in planet, it is not surprising that we obtain contradicting results for the ages from these parameters.

Although the radius determination using Gaia parallax is expected to be more accurate, we also determined the $\rm log(g)$ using Ca\,I, Mg\,I and Na\,I as an independent test.
Determining the radius of a star from log(g) is less accurate then either using the stellar density derived from the light-curve fit, or the
radius determined by combining the distance of the star combined with the SED. 
However, comparing the stellar density obtained from the spectral analysis with that obtained from the light-curve fitting of the star is a good test of the spectral analysis
\citep{2012A&A...537A.136G}. 
The results of these tests are discussed in Section\,\ref{sectIIa}.
Using Ca\,I we derive $\rm log(g)=4.20\pm0.06$, using  Mg\,I $\rm log(g)=4.26\pm0.08$, and using Na\,I $\rm log(g)=4.2\pm0.04$. Given the temperature of this star, the value derived from the Mg\,I is the most accurate. 
We obtain the element abundances of iron $\rm [Fe/H] = -0.03\pm0.05$ (dex), calcium $\rm [Ca/H] = -0.03\pm0.05$ (dex), sodium $\rm [Na/H] = 0.00\pm0.05$ (dex), and magnesium $\rm [Mg/H] = -0.03\pm0.05$ (dex).

Using $T_{eff}$, log(g) and [Fe/H] derived spectroscopically, the Bayesian estimation of stellar parameters, and the MESA isochrones \citep{2006AuA...458..609D,2014MNRAS.445.2758R,2017MNRAS.467.1433R} we obtain $\rm R_{*}=1.007_{-0.024}^{+0.050}$ $\rm R_\odot$, and $\rm M_{*}=0.886_{-0.014}^{+0.014}$ $\rm M_\odot$.
The mass and radius of the star derived 
spectroscopically and using SED-fitting are independent from each other, since we used $T_{eff}$ and [Fe/H] only as priors for the SED fitting.

We also made a second analysis of the ESPRESSO spectra. 
Adding up all ESPRESSO spectra, weighted by their signal-to-noise ratio, we obtain a spectrum with a S/N of 228. We then obtained the equivalent width of the FeI and FeII lines from the ispec framework \citep{Blanco14,Blanco19}. The equivalent width are the fitted the ATLAS (LTE) model atmospheres \citep{Kurucz05}, SYNTHE suite stellar atmosphere modeling code and WIDTH radiative transfer code to derive chemical abundances \citep{Sbordone04}. 

From this analysis we obtain $\rm T_{eff}= 5488\pm60$\,K,
$\rm log(g)= 4.28\pm0.2$ and $\rm [FeH]=0.04\pm0.08$. Converting again these
values into the mass, radius and age of the star gives $\rm M_{*}=0.943\pm0.029\,M\odot$, and $\rm R_{*}=0.983\pm0.013\,R\odot$ and $7.3\pm2.0$\,Gyrs. Using these parameters gives a distance of $\rm d=244.6\pm2.9$\,pc in perfect agreement with the distance determined by Gaia. Both mass and radius values 
derived in this section are shown as a red points in Figure \ref{mass_radius_star}. 

\subsection{Comparing the new mass and radius determination of the host star with previously determinations}
\label{sectIIa}

As mentioned above, the mass, radius and temperature of the host star has already been determined previously
\citep{2017AJ....153...82A,2017AJ....153..271S,2017AuA...608A..93G, 2019ApJ...883...79D,2021PSJ.....2..152A,  2022AuA...658A.132S,2023AJ....165...97R}. How do the new values compare to the previous determinations?
The mass, radius,  $\rm T_\mathrm{eff}$ and, $\rm [Fe/H]$ from the literature and derived in this work are given Table\,\ref{tab:starI}.

In Figure\,\ref{mass_radius_star} we compare the mass and radius measurements obtained for the host star from the literature (black points individual measurements, blue triangle average),  with the spectroscopic determination (red points), and with the values obtained using ARIADNE and ISOCHRONES (green points).

Since ARIADNE and ISOCHRONES take advantage of the accurate distance determination obtained in Gaia DR3,
these values are the preferred ones. Some of the previous determinations are based on Gaia DR1, or Gaia DR2, and thus have larger errors compared Gaia DR3. 

There is another possibility to verify the mass and radius derived for the star. 
As explained in Section \ref{sectIVa}, the density of the star can also be derived from light-curve fitting.
Comparing the stellar density derived from the light-curve and stellar modelling thus is an
excellent test.

\subsection{The radii of the planets in respect to the radius of the host star}

\subsubsection{Analysis of the K2 and TESS light curves}
\label{sectIIIc}

As the name already indicates, K2-106 was originally found in the K2 survey. 
136 transits were obtained during  the K2-mission.
Up to now, only 12 transits have been observed with the TESS satellite. 
The cadence time of K2 data is 30 minutes. 
We super-sampled the transit model by a factor of 10 to account for the K2 long-cadence data as described in 
\citep{2010MNRAS.408.1758K,2019MNRAS.482.1017B}.
The quality of the TESS light-curves is much lower than those obtained in the K2 mission. 
We therefore use only the K2 data for the radius determination, and 
the K2 together with the TESS for the ephemeris.
There are several ways how to extract the light-curves, how to remove the instrumental effects and how to remove stellar activity. The star is, however, quite inactive.
We tried out the light curves provided by \citet{2014PASP..126..948V}, and those obtained with the K2SC algorithm \citep{2016MNRAS.459.2408A}
\footnote{https://archive.stsci.edu/prepds/k2sc/}. 

Figure \ref{RadiusRatio} shows the values obtained for the two light curves with, and without using the stellar density derived as an informative prior (See Section \ref{sectIVa} for details).  
Both light-curves gave consistent results, but the errors are smaller for the K2SC data. 

\begin{figure}
\includegraphics[height=0.37\textwidth,angle=-90.0]{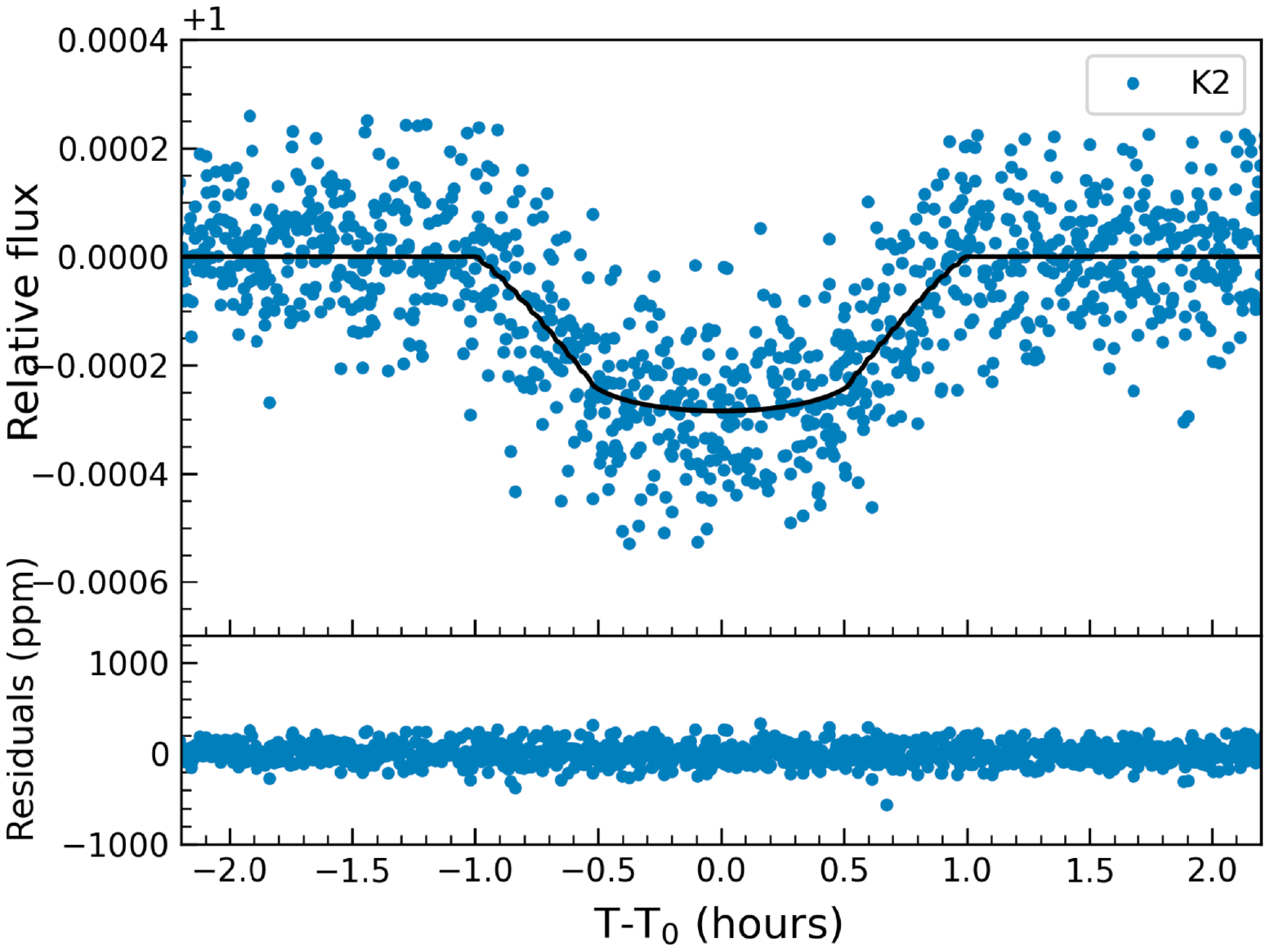}
\caption{Phase folded light-curve obtained with K2 of K2-106\,b, after removing a few outlier using a 3 $\sigma$ clipping.}
\label{photI}
\end{figure}

\begin{figure}
\includegraphics[height=0.37\textwidth,angle=-90.0]{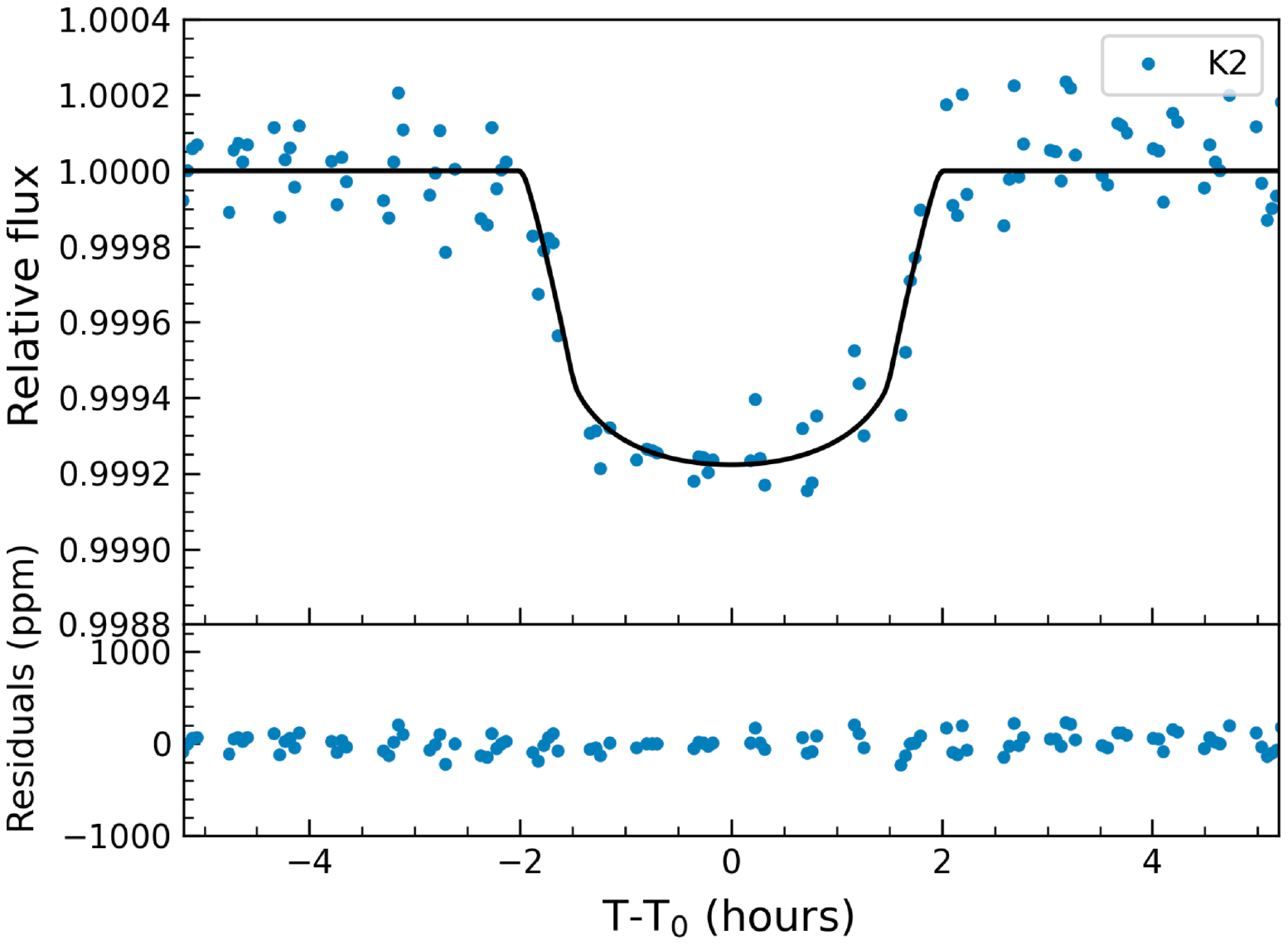}
\caption{Phase folded light-curve obtained with K2 of K2-106\,c
after removing a few outlier using a 3 $\sigma$ clipping.}
\label{photII}
\end{figure}

As described in detail in Section \ref{sectIVa}, we model the light-curves and the RV curve together. 
We included a photometric jitter term in the fit to account for
instrumental noise that is not included in the nominal error bars.
The photometric jitter term for K2 is $\rm \sigma_\mathrm{K2}=0.0000882\pm0.0000018$.
Figure \ref{photI} and Figure \ref{photII} show the fit to the phase folded light-curves using the
combined model. Outliers were removed using 3 $\sigma$ clipping criterion.

The RV-measurements are discussed in  Section \ref{sectIIId}. The off-sets between different instruments 
and their jitter terms are given in Table\,\ref{tab:RV}, and Table\,\ref{tab:jitter}.

\subsubsection{Comparing the ratio of the radius of the planets to the host star with previous determinations}
\label{sectIIb}

\begin{figure}
\includegraphics[height=0.35\textwidth,angle=0.0]{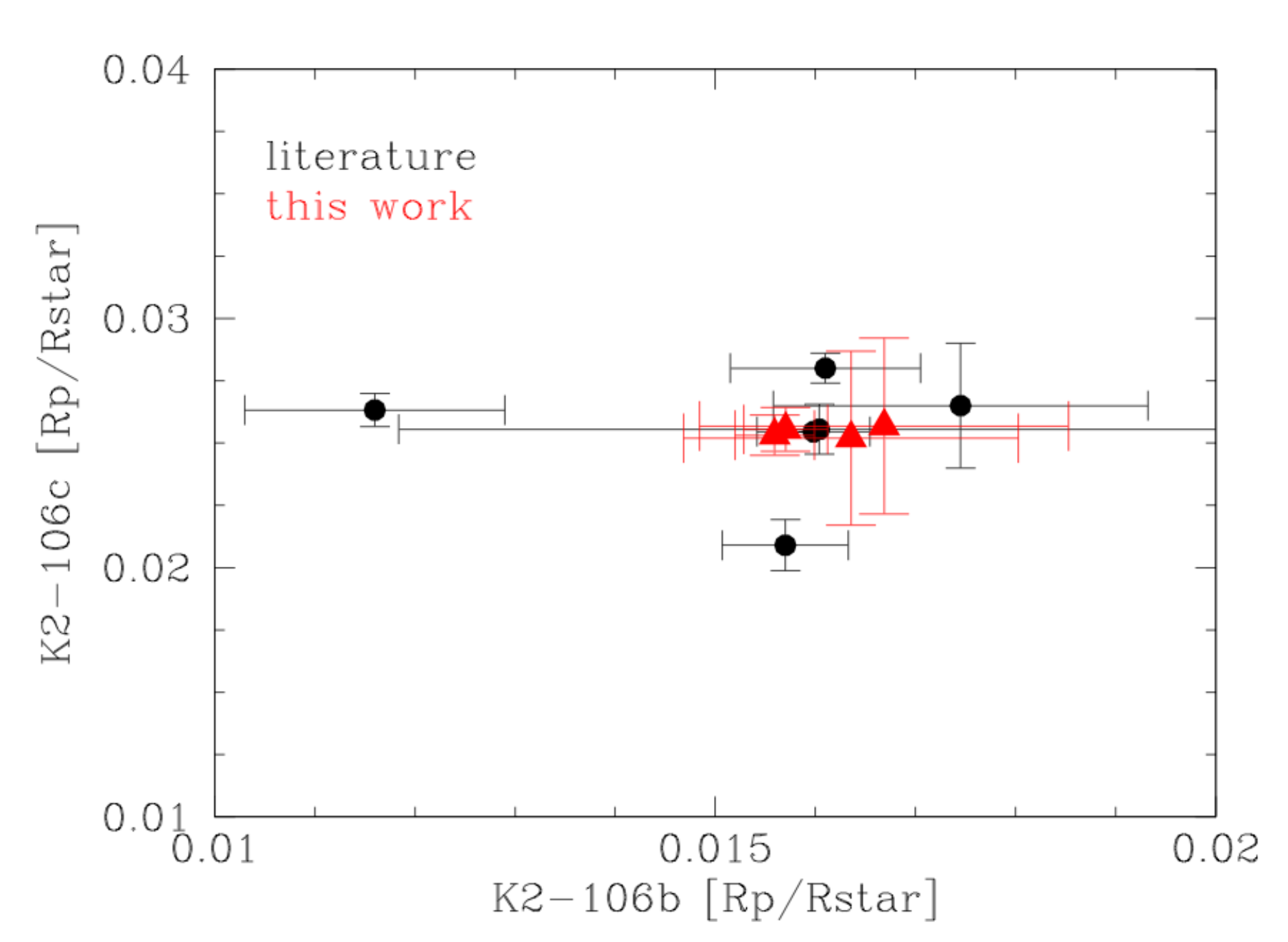}
\caption{Ratio of the radius of the planets K2-106\,b and K2-106\,c to the radius of the
host star. Values from the literature (black), and the vales derived in this work (red triangle).}
\label{RadiusRatio}
\end{figure}

Figure \ref{RadiusRatio} shows the previous determinations of the ratio of the radii of the planets to the radius of the host star. The black points are values taken from the literature \citep{2017AuA...608A..93G,2017AJ....153...82A,2017AJ....153..271S,2019ApJ...883...79D,2022AuA...658A.132S, 2023AJ....165...97R}.  
Since \citet{2019ApJ...883...79D} and \citet{2022AuA...658A.132S} did not publish the
values for K2-106\,c we simply used the 
average $\rm Rp/R_{star}$-values for K2-106\,c from the literature to show
the values for K2-106\,b in this figure. 
The ratio of the radius of K2-106\,b to the host star is within the errors the same.
Using the variance as the error it is $R_{K2-106b}/R_{star}=0.0160\pm0.0006$, and $\rm R_{K2-106c}/R_{star}=0.0270\pm0.0009$, respectively. 
We also show the radius-ratios derived in Section \ref{sectIIIc} as red points. The new determinations of the ratios of the radii of the planets to the host stars are in line with the previous determinations.

\subsection{RV measurements obtained with ESPRESSO combined with previous measurements}
\label{sectIIId}

We obtained 23 new RV-measurements of K2-106 with the ESPRESSO spectrograph that were reduced and extracted using the dedicated ESPRESSO pipeline
\citep{2021A&A...645A..96P}. The RVs were determined by using a cross-correlation method with a numerical mask that corresponds to a G8 star.
The RVs were obtained in the usual manner by fitting a Gaussian function to the average cross-correlation function (CCF) \citep{1996A&AS..119..373B, 2021A&A...645A..96P}.
The data reduction pipelines of this instrument also provides the absolute RV. 
The median error for the ESPRESSO data is 1.7 $\rm ms^{-1}$.
For comparison, the 32 HARPS spectra that we used previously had a median error of 3.2 $\rm ms^{-1}$, the 13 PSF spectra 3.0 $\rm ms^{-1}$, the three HDS spectra 5.0 $\rm m\,s^{-1}$,  and the 6 FIES spectra 4.8 $\rm ms^{-1}$ \citep{2017AuA...608A..93G}.

Because our main interest is the mass determination of the inner planet K2-106\,b,
we decided to take several spectra per night when possible.
The RV-values obtained with ESPRESSO are listed in Table\,\ref{tab:RV}.

\begin{figure}
\includegraphics[height=0.27\textwidth,angle=0.0]{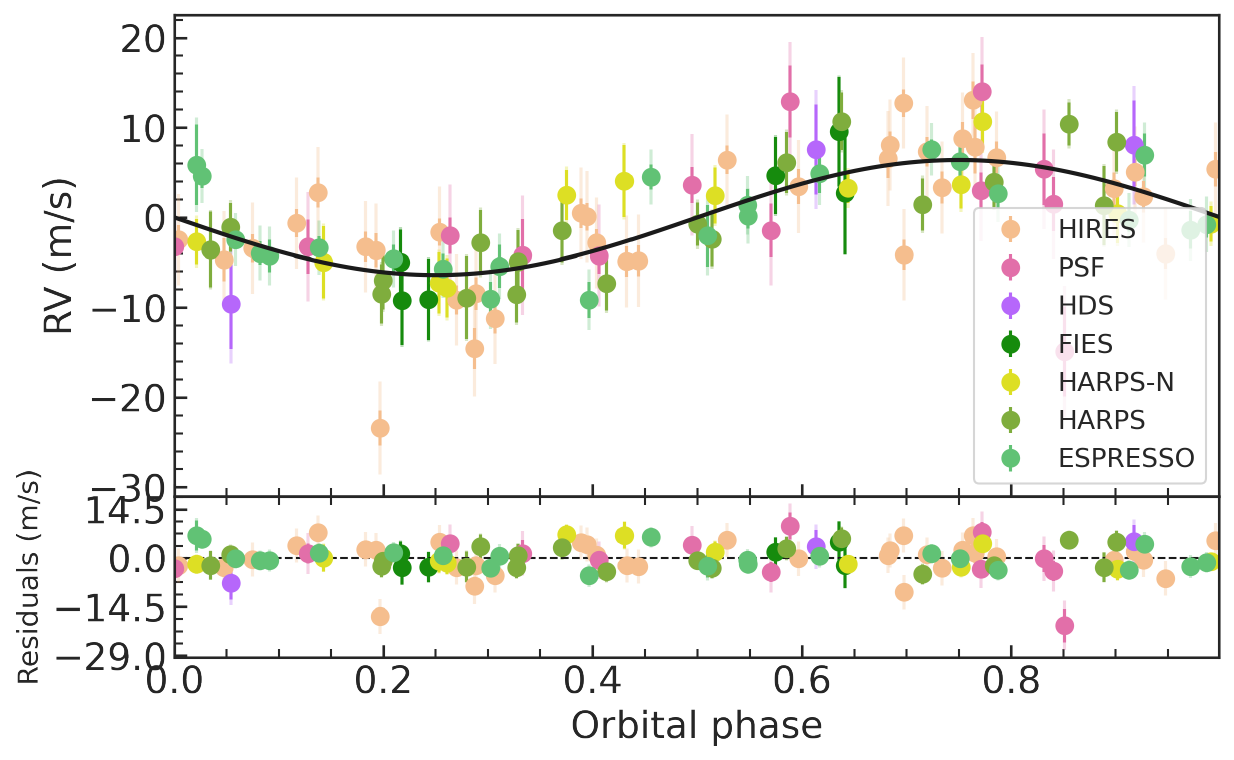}
\caption{Phase-folded RV-curve of K2-106\,b.}
\label{RV1}
\end{figure}

\begin{figure}
\includegraphics[height=0.27\textwidth,angle=0.0]{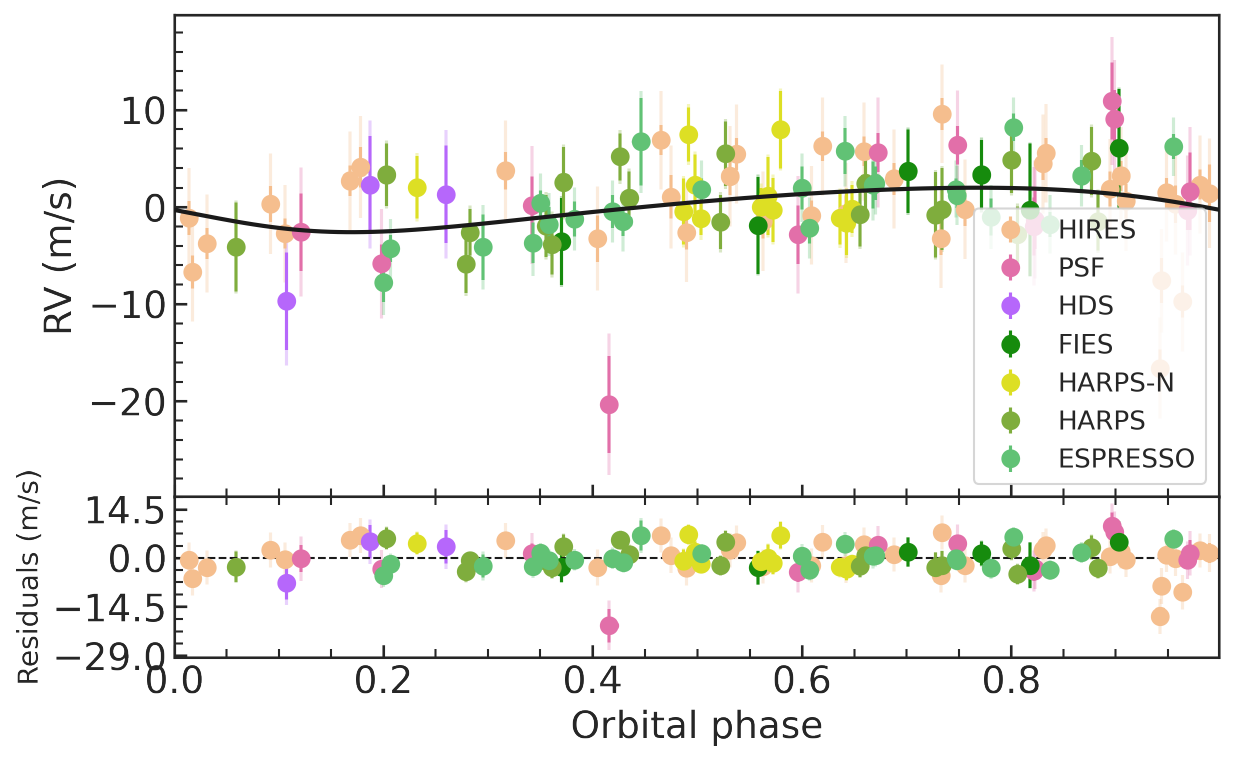}
\caption{Phase-folded RV-curve of K2-106\,c.}
\label{RV2}
\end{figure}

In order to combine the RVs obtained previously with the ESPRESSO data, we determined instrumental off-sets between the instruments and the jitter-terms. The off-sets and jitter-terms are listed in Table\,\ref{tab:jitter}. 
The accuracy of the RVs obtained with ESPRESSO is higher than that of the other instruments but the high jitter term indicates that the star was a bit more active during ESPRESSO
observations.
Figures \ref{RV1},\ref{RV2} show the phase-folded RV-curves of K2-106b and K2-106c, respectively. 

\begin{table*}
\caption{Radial velocities obtained with ESPRESSO}
\begin{tabular}{c c c c c}
\hline
\noalign{\smallskip}
$\rm BJD_{TDB}$     & RV     & FWHM   & BIS    & S/N \\
-2 450 000  & [$\rm km\,s^{-1}$] &  [$\rm km\,s^{-1}$] & [$\rm km\,s^{-1}$] &  \\
\hline
8703.75031525 & $-15.8276\pm0.0034$ & $7.46347\pm0.0067$ & $-0.0931\pm0.0067$ & 29.4 \\
8704.91475462 & $-15.8232\pm0.0018$ & $7.48160\pm0.0038$ & $-0.0952\pm0.0037$ & 43.5 \\
8705.75671620 & $-15.8180\pm0.0045$ & $7.49173\pm0.0089$ & $-0.0965\pm0.0089$ & 24.7 \\
8707.81075258 & $-15.8176\pm0.0022$ & $7.48481\pm0.0044$ & $-0.0923\pm0.0044$ & 38.9 \\
8707.90836233 & $-15.8198\pm0.0016$ & $7.48639\pm0.0032$ & $-0.0876\pm0.0032$ & 47.6 \\
8708.74798097 & $-15.8278\pm0.0017$ & $7.48843\pm0.0035$ & $-0.0877\pm0.0035$ & 45.4 \\
8709.77704796 & $-15.8243\pm0.0011$ & $7.48605\pm0.0023$ & $-0.0883\pm0.0023$ & 61.2 \\
8709.79036694 & $-15.8258\pm0.0013$ & $7.48668\pm0.0025$ & $-0.0921\pm0.0025$ & 56.6 \\
8717.72563374 & $-15.8264\pm0.0021$ & $7.47178\pm0.0042$ & $-0.0896\pm0.0042$ & 40.1 \\
8717.82099476 & $-15.8283\pm0.0013$ & $7.48843\pm0.0026$ & $-0.0926\pm0.0026$ & 54.5 \\
8718.74229989 & $-15.8179\pm0.0017$ & $7.48184\pm0.0033$ & $-0.0882\pm0.0033$ & 46.8 \\
8718.87715170 & $-15.8248\pm0.0015$ & $7.48387\pm0.0030$ & $-0.0945\pm0.0030$ & 50.1 \\
8719.86954645 & $-15.8157\pm0.0011$ & $7.48672\pm0.0021$ & $-0.0907\pm0.0021$ & 62.3 \\
8721.69988499 & $-15.8153\pm0.0024$ & $7.50070\pm0.0048$ & $-0.0895\pm0.0048$ & 36.7 \\
8724.71755081 & $-15.8268\pm0.0016$ & $7.48293\pm0.0033$ & $-0.0960\pm0.0033$ & 47.2 \\
8777.68002329 & $-15.8223\pm0.0013$ & $7.48530\pm0.0026$ & $-0.0950\pm0.0026$ & 50.5 \\
8782.52691959 & $-15.8355\pm0.0020$ & $7.47913\pm0.0040$ & $-0.0942\pm0.0040$ & 36.8 \\
8782.61367352 & $-15.8262\pm0.0014$ & $7.49631\pm0.0028$ & $-0.0957\pm0.0028$ & 46.9 \\
8784.63761295 & $-15.8291\pm0.0018$ & $7.48978\pm0.0036$ & $-0.0915\pm0.0036$ & 54.6 \\
8788.76247468 & $-15.8275\pm0.0024$ & $7.50070\pm0.0048$ & $-0.0898\pm0.0048$ & 32.4 \\
 \\
\hline
\end{tabular}
\label{tab:RV}
\end{table*}

\begin{table}
\caption{Off-sets and jitter-terms}
\begin{tabular}{l r l}
\hline
\noalign{\smallskip}
Instrument & off-set & jitter \\
 & $\rm [m\,s^{-1}]$ & $\rm [m\,s^{-1}]$ \\
\hline
HIRES     & $-22.22_{-0.88}^{+0.88}$ & $ 4.79_{-0.60}^{+0.74}$ \\
PSF       & $0.63_{-1.85}^{+1.77}$  & $5.36_{-1.68}^{+2.07}$  \\
HDS       & $24.28_{-4.57}^{+4.84}$  & $ 3.94_{-3.21}^{+9.48}$  \\
FIES      & $98.92_{-2.36}^{+2.32}$  & $ 1.24_{-0.95}^{+2.48}$ \\
HARPS-N   & $-15736.08_{-1.10}^{+1.09}$ & $ 1.37_{-1.01}^{+1.48}$ \\
HARPS:    & $-15732.63_{-0.85}^{+0.83}$ & $ 1.31_{-0.95}^{+1.21}$ \\
ESPRESSO: & $-15823.74_{-0.73}^{+0.73}$ & $ 2.74_{-0.56}^{+0.70}$  \\
\hline
\end{tabular}
\label{tab:jitter}
\end{table}

\subsection{Radii and masses of the two planets}
\label{sectIVa}

The mass and radius of the host star are key factors for the determination of the masses and radii of planets. 
As explained in Sections \ref{sectIIa}, \ref{sectIIIa}, and \ref{sectIIIb}, we 
obtained six different sets of stellar parameters. 
However, if we do not count 2MASS-Gaia DR3 method, because it gives only the radius of the star, and the value derived directly from the log(g) and $\rm R_*$, we have obtained 
four new values for the mass and radius of the star.

We determined the masses and radii of the planets using the PYANETI-code \citep{2019MNRAS.482.1017B,2022MNRAS.509..866B}.
PYANETI performs a multiplanet radial velocity and transit data fitting. 
The code uses a Bayesian approach combined with a Markov chain Monte Carlo sampling to estimate the parameters of planetary systems.
We added a photometric and an RV jitter term to account for instrumental noise not included in the nominal uncertainties.
We used the standard set up previously used in other articles
\citep{2019MNRAS.490..698B,2022MNRAS.514.1606B,2023MNRAS.522.3458B,2021MNRAS.505.4684G,2022NatAs...6..736S,2022A&A...666A.184P}. 
A good sampling is assured using a number of chains which is at least as twice as the amount of parameters. 
We sampled the parameter space using 500 Markov chains. We created the posterior distributions using the last 5000 iterations of the converged chains with a thin factor of 10. 
We used the convergence test developed by \cite{1992StaSc...7..457G}, as described in \cite{2019MNRAS.490..698B}. This approach leads to a distribution of 250 000 points for each model parameter per distribution. The best estimates and their 1-$\sigma$ uncertainties were taken as the median and the 68\% limits of the credible interval of the posterior distributions. \footnote{In \cite{2019MNRAS.490..698B}, they defined convergence as when chains have a scaled potential factor \^{R} = \begin{math}\sqrt{[W(n-1)/n + B/n]/W} < 1.02\end{math} for all the parameters \citep{2004Gelman}, where B is the 'between-chain' variance, W is the 'within-chain' variance, and n is the length of each chain.}

We obtained the density of the star from the light-curve without using stellar density
as informative prior and also using the stellar density as an uninformative prior. 
We did not find any significant difference between the two. 
Using the stellar density obtained with ISOCHRONES as a informative prior we obtain from the light-curve modelling of the inner planet a stellar density of  $\rm \rho_{star}=1.349_{-0.089}^{+0.089}\,g\,cm^{-3}$.
We did the same analysis using the stellar density from ARIADNE as informative prior. 
In this case we obtain $\rm \rho_{star}=1.348_{-0.106}^{+0.116}\,g\,cm^{-3}$. 
The difference is insignificant but the errors are larger if we use the ARIADNE values.

We can also compare the stellar density derived from the light-curve with 
the stellar density derived from the four methods described in Sections \ref{sectIIIb} and \ref{sectIIIa}.
This is possible, because the density of the star can be deviled from the light-curve without knowing the
mass and radius of the host star \citep{2017AJ....154..228S}. We find that the difference is smallest 
for the stellar parameters obtained using ISOCHRONES. 
The stellar density derived from the mass and radius using ISOCHRONES is
$\rm \rho_{star}=1.327_{-0.058}^{+0.060}\,g\,cm^{-3}$. 
With ARIADNE we obtain $\rm \rho_{star}=1.315_{-0.106}^{+0.116}\,g\,cm^{-3}$.
In Section \ref{sectIIId} we also derived the mass and radius of the star using the stellar parameters
derived from the ESPRESSO spectra. Using these values we obtain densities of $1.22^{+0.17}_{-0.09}$ $g\,cm^{-3}$ and $1.40\pm0.07$ $g\,cm^{-3}$ for the star, respectively.

The difference between the density derived from the four sets of stellar parameters, and
from the light-curve fitting are thus small, especially for the stellar model derived from ISOCHRONES. 
The mass and radius derived from ISOCHRONES matches also the average mass and radius values of the star from the literature best. Because the errors for the stellar model from ISOCHRONES are also smaller, we adopt these values. However, we will still discuss in Section\,\ref{sectV} if it makes a difference if we
use one of the other sets of stellar parameters. 

The difference between stellar density from ISOCHRONES as an informative prior or as an uninformative prior
are insignificant. If we use stellar density from ISOCHRONES as an informative prior or as an uninformative prior, we find for the inner planet:
$\rm Rp=1.780_{-0.061}^{+0.065}\,R_{\oplus}$ versus
$\rm Rp=1.767_{-0.060}^{+0.062}\,R_{\oplus}$ and
$\rm Mp=8.61_{-0.83}^{+0.84}\,M_{\oplus}$ versus
$\rm Mp=8.58_{-0.80}^{+0.83}\,M_{\oplus}$.

The results obtained for the two planets are listed in Table\,\ref{tab:planet}.  
The phase-folded light-curves are shown in Figures\,\ref{photI}, and \ref{photII}. 
The RV-curves are displayed in Figures\,\ref{RV1}, and \ref{RV2}.
The posterior distributions of the parameters are shown in Figure\,\ref{fig:posteriors}, and \ref{fig:posteriorsB}.

\begin{table*}
\caption{Parameters K2-106\,b and K2-106\,c}
\begin{tabular}{l l l}
\hline
\noalign{\smallskip}
Parameter & K2-106\,b & K2-106\,c \\
\hline
$\rm T_0$ & $2457394.00907_{-0.00069}^{+0.00067}$ & 
$2457405.7320_{-0.0016}^{+0.0016}$ \\
P [days]  & $ 0.5713127\,\pm\,0.0000055$ & $13.33989_{-0.00070}^{+0.00068}$ \\
e         & 0.0 & $0.17_{-0.11}^{+0.11}$ \\
b         & $0.24_{-0.14}^{+0.11}$ & $ 0.53_{-0.20}^{+0.13}$ \\
$\rm a/R_{*}$      & $2.855_{-0.065}^{+0.065}$ & $ 23.32_{-0.53}^{+0.50}$ \\
$\rm Rp/R_{*}$, K2 & $0.01553_{-0.00028}^{+0.00029}$ & $0.02659_{-0.00073}^{+0.00080}$ \\
i\,[deg]           & $85.2_{-2.4}^{+2.9}$ & $ 88.62_{-0.17}^{+0.40}$ \\  
$\rm a/R_{*}$, K2  & $2.855_{-0.065}^{+0.061}$ & $23.32_{-0.52}^{+0.50}$ \\
a\,[AU]            & $0.01314_{-0.00034}^{+0.00032}$ & $0.1073_{-0.0027}^{+0.0026}$ \\
$\rm depth, K2\,[ppm]$ & $241.0_{-8.7}^{+9.0}$ & $ 693_{-36}^{+47}$  \\
$\rm T_{tot} [h]$      & $1.540_{-0.031}^{+0.033}$ & $3.661_{-0.073}^{+0.078}$  \\
$\rm T_{full} [h]$     & $1.487_{-0.033}^{+0.034}$ & $3.389_{-0.085}^{+0.075}$  \\
$\rm T_{in/eg} [h]$    & $0.026_{-0.001}^{+0.002}$ & $0.130_{-0.028}^{+0.041}$ \\
$\rm K\,[m\,s^{-1}]$   & $6.36_{-0.57}^{+0.57}$ & $2.14_{-0.69}^{+0.74}$ \\
$\rm M_{p}\,[M_{\oplus}]$    & $7.80_{-0.70}^{+0.71}$    & $7.32_{-2.38}^{+2.49}$ \\
$\rm R_{p},K2\,[R_{\oplus}]$ & $1.676\,\pm\,0.037$ & $2.84_{-0.08}^{+0.10}$ \\
$\rm \rho_{p}\,[g\,cm^{-3}]$ & $9.09\,\pm\,0.98$ & $1.72_{-0.58}^{+0.66}$ \\
$\rm T_{eq}^{(1)}\,[K]$        & $2299_{-35}^{+36}$ & $804_{-12}^{+13}$ \\
$\rm \lambda^{(2)}$            & $15.3_{-1.4}^{+1.5}$ & $24_{-8}^{+8}$ \\
$\rm Insolation\,[F_{\oplus}]$ & $4656_{-274}^{+300}$ & $70_{-4}^{+5}$  \\
\hline
\end{tabular}
\\
$^{(1)}$ {\begin{math}\rm T_\mathrm{eq} = T_\mathrm{eff}\sqrt{\frac{R_\star}{2a}}(1-A_B)^{1/4}\end{math}, using an albedo $A_{B}$=0}; 
$^{(2)}\lambda$: Jeans escape parameter \citep{2017AuA...598A..90F}. 
\label{tab:planet}
\end{table*}

\section{Discussion}
\label{sectV}

\begin{figure}
\includegraphics[height=0.35\textwidth,angle=0.0]{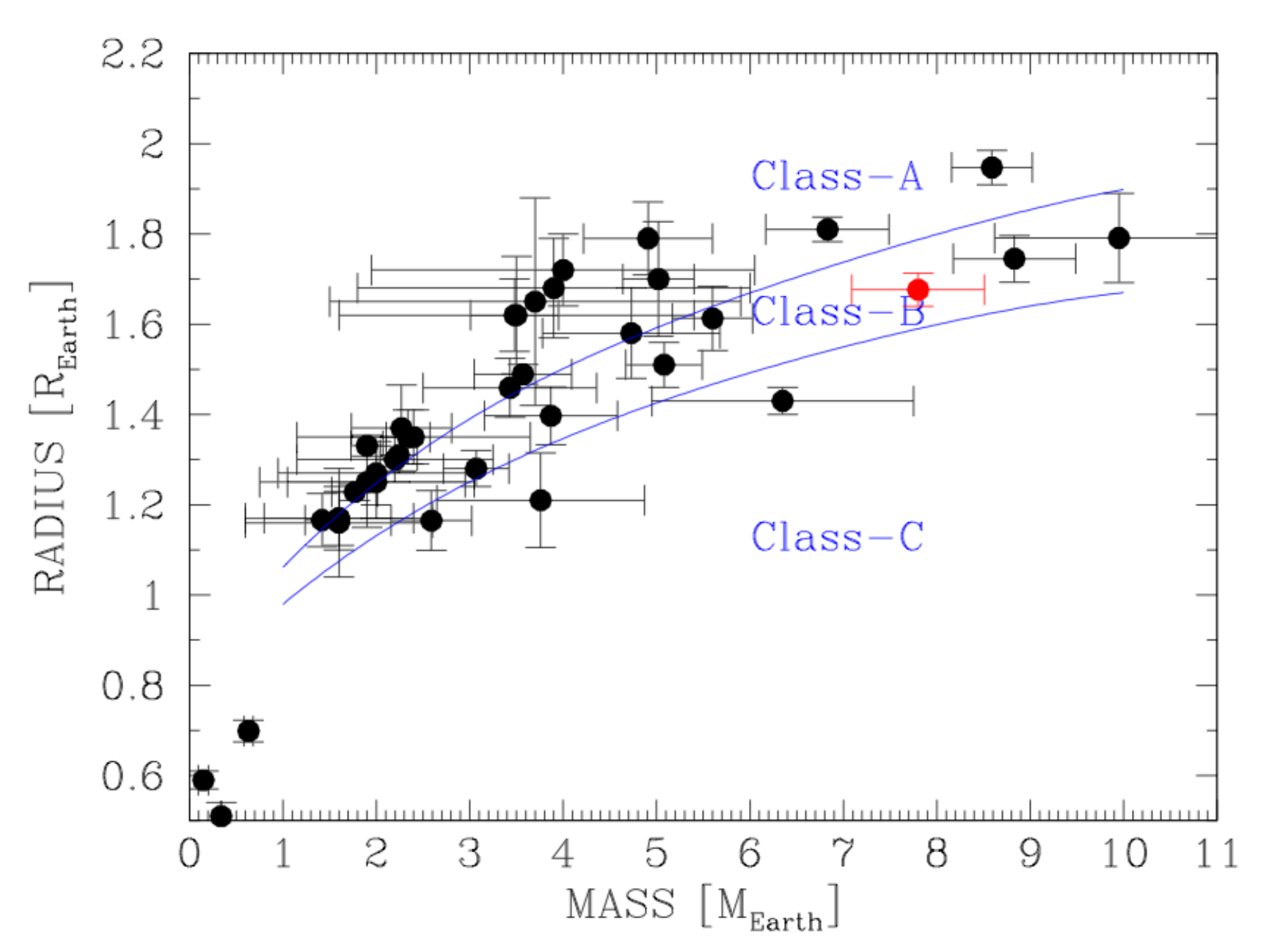}
\caption{Mass-Radius diagram for USPs. The red point is K2-106\,b. 
The dark blue lines are the lower and upper limits for planets 
an Earth-like core radius fraction calculated by \citep{2018Icar..313...61H}. 
The upper line represents a core which is composed of 80\% iron and 20\% other elements,
and a mantle that is made of $\rm MgSiO_3$. The lower curve is a pure Fe core with a $\rm FeSiO_3$ mantle.}
\label{USPsMpMr}
\end{figure}

\begin{figure}
\includegraphics[height=0.35\textwidth,angle=0.0]{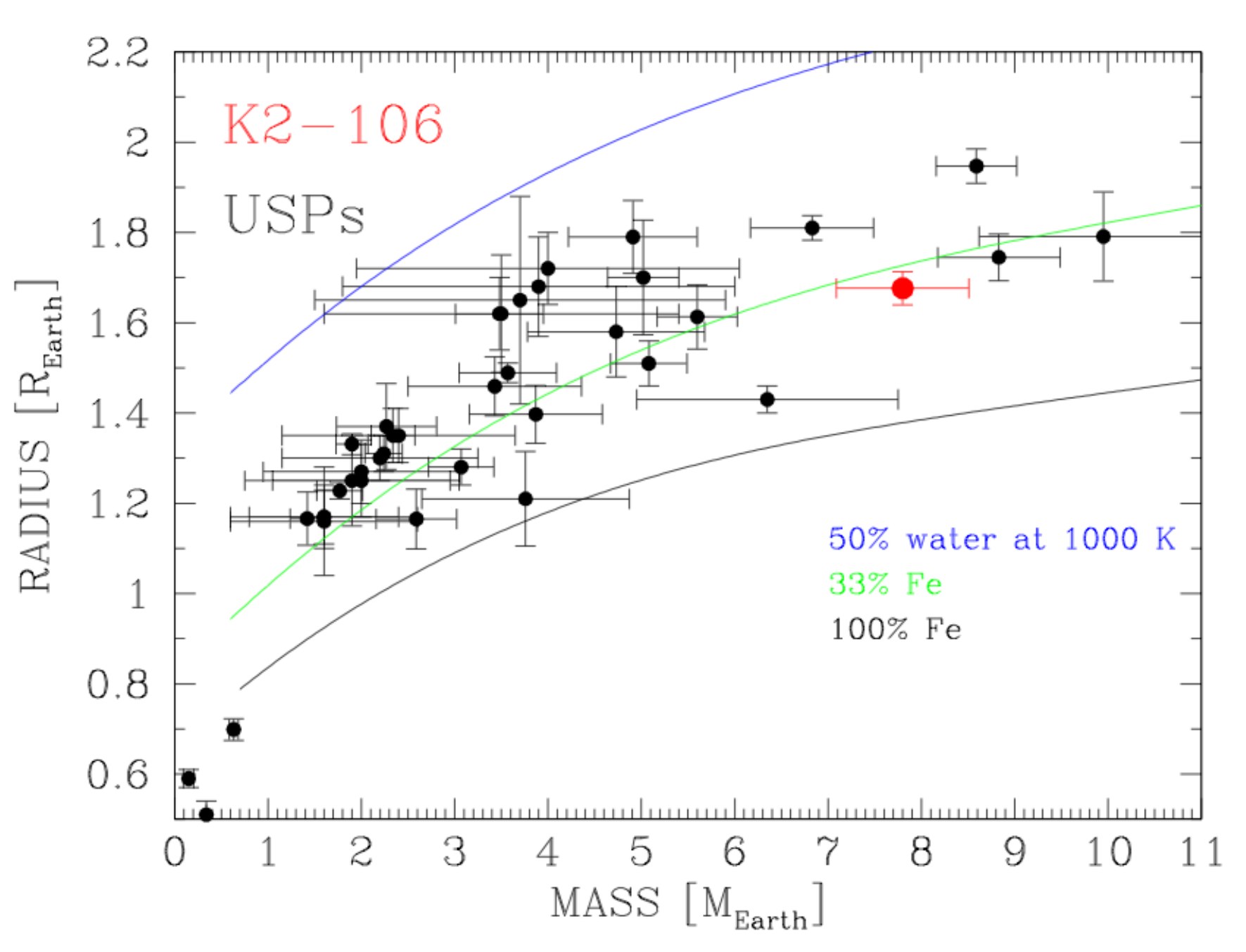}
\caption{Classical mass-radius diagram for USPs. The red point is K2-106\,b, the other points are the
other USPs listed in Table\,\ref{tab:USPs}. The mass-radius compositions
for 100\% iron (black line), 33\% iron (green line) and 50\% water at 1000\,K (blue line) are taken from \citet{2019PNAS..116.9723Z}.}
\label{Fig12}
\end{figure}

The K2-106 system is one of the best systems for measuring the densities of low-mass planets, but what precision has been achieved?

The first error source is the radius and the mass of the host star. 
Some of the previous studies of this system have used older versions of the Gaia measurements.
The Gaia DR3 measurements of the radius significantly improves the accuracy. 
We have determined the radius and the mass of the  star using four different methods. The values from the literature and our new values are given in Table\,\ref{tab:starI}. 

The density of the inner planet derived with ARIADNE is 
$\rm 8.25^{+1.15}_{-1.02}$ $\rm g\,cm^{-3}$. With ISOCHRONES we obtain $\rm 9.09_{-0.98}^{+0.98}$ $\rm g\,cm^{-3}$.
Within the errors, the two values are the same. 
Taking the variance of the two values, the precision with which the
density of the inner planet could be determined is 6.9\%.
The density of K2-106\,b has now been determined to a higher precision than for most other low-mass USPs. 

If we want to asses the nature of a planet we also have to take the errors of the theoretical models in to account.
\citet{2018Icar..313...61H} argue that it is not known what the exact composition of the core and mantle of an exoplanet is.
They calculate the mass-radius diagram for planets with a Mercury-like, Earth-like, and Moon-like
core-to-mantle fraction using four different mantle and six different core compositions. 
A planet with the highest density has pure Fe core and a $\rm Fe_2SiO_3$ mantle. 
A rocky planet with the lowest density has a core that contains 80\% iron, and 20\% Aluminium and other light elements, and a $\rm MgSiO_3$ mantle.
In this model, planets with an Earth-like core-to-mantle ratio thus can have different densities, depending on the exact composition of the core and the mantle.
We define as a class-B planet, a planet that has the same core-to-mantle ratio as the Earth allowing for different compositions of the core and the mantle.  

Fig.\ref{USPsMpMr} shows the mass-radius measurements for known USPs including K2-106\,b. The lines indicate the maximum and minimum radius for planets with the core-to-mantle ratios as the Earth.
Fig.\ref{Fig12} shows the classical mass-radius diagram for USPs using the  models published by \citet{2019PNAS..116.9723Z}. 
Many USPs fall into the regime of Class-B planets, but there are also many that are in Class-A.

K2-106\,b is in Class-B, but this does not mean that K2-106\,b must be Earth-like. 
Even if it would have a similar composition as the Earth, 
an USPs is always unlike the Earth.
For example, USPs have lava oceans, because of the radiative heating by the host star
\citep{2010ASPC..430..409B, 2010ApJ...709L..95B}. Furthermore, USPs are not only
radiatively heated but also tidally, magnetically, by flares and by CMEs from the
host star \citep{2017NatAs...1..878K,2020A&A...644A.165B,2021A&A...653A.112L,2022ApJ...941L...7G}.
It is thus plausible that many lmUSPs have lava oceans and thus outgassed, or hybrid atmospheres.
Such an atmosphere would then contain heavier, non-volatile elements.
For example, \citet{2022PSJ.....3...93B} showed that the atmospheres of planets with lava oceans should be carbon-rich. 
This means that although K2-106\,b is an class Class-B planet, it does not have to have the same structure, and composition as the Earth. 
For example, K2-106\,b could have a core that is larger than that of the Earth, and an atmosphere.
At the present stage, we do not even know whether K2-106\,b is simply a bare rock, or a planet with a lava-ocean and an atmosphere. 

Thus, additional observations are needed to find out what kind of a planet it is. First of all, we need to find out, whether it has a lava ocean and an atmosphere, or not.
\citet{2022A&A...661A.126Z} have studied the observability of evaporating lava worlds, and
\citet{2015ApJ...801..144I} have calculated the spectrum of an atmosphere composed of gas-species from a magma ocean.  
Phase curves  of K2-106\,b obtained with the JWST would allow to find out if there is 
a lava ocean and an atmosphere. Transit observations obtained with the JWST, or with CRIRES$\rm ^{+}$ would also allow us to find out what the composition of the atmosphere is, if the planet has one.
 
However, using a three-layer interior structure mode \citet{2018MNRAS.476.2613S} have shown that 
planets with very different compositions can have the same bulk density. 
Thus, even if K2-106b has no atmosphere, this does not mean that it must have an Earth-like composition. There are still other possibilities.
Observations alone can not rule out all other possibilities, 
and an improved formation theory is need to narrow down the possibilities.

To put K2-106\,b into perspective, Table\,\ref{tab:USPs} and Figure \ref{USPsMpMr}
show all known lmUSPs with radius- and mass-determination. 
The dark blue lines in Figure \ref{USPsMpMr} are the lower and upper limits for planets with Earth-like composition from \citet{2018Icar..313...61H}.
For comparison we also show the classical mass-radius diagram using the models from \citet{2019PNAS..116.9723Z} in Figure \ref{Fig12}.

\begin{figure}
\includegraphics[height=0.30\textwidth,angle=0.0]{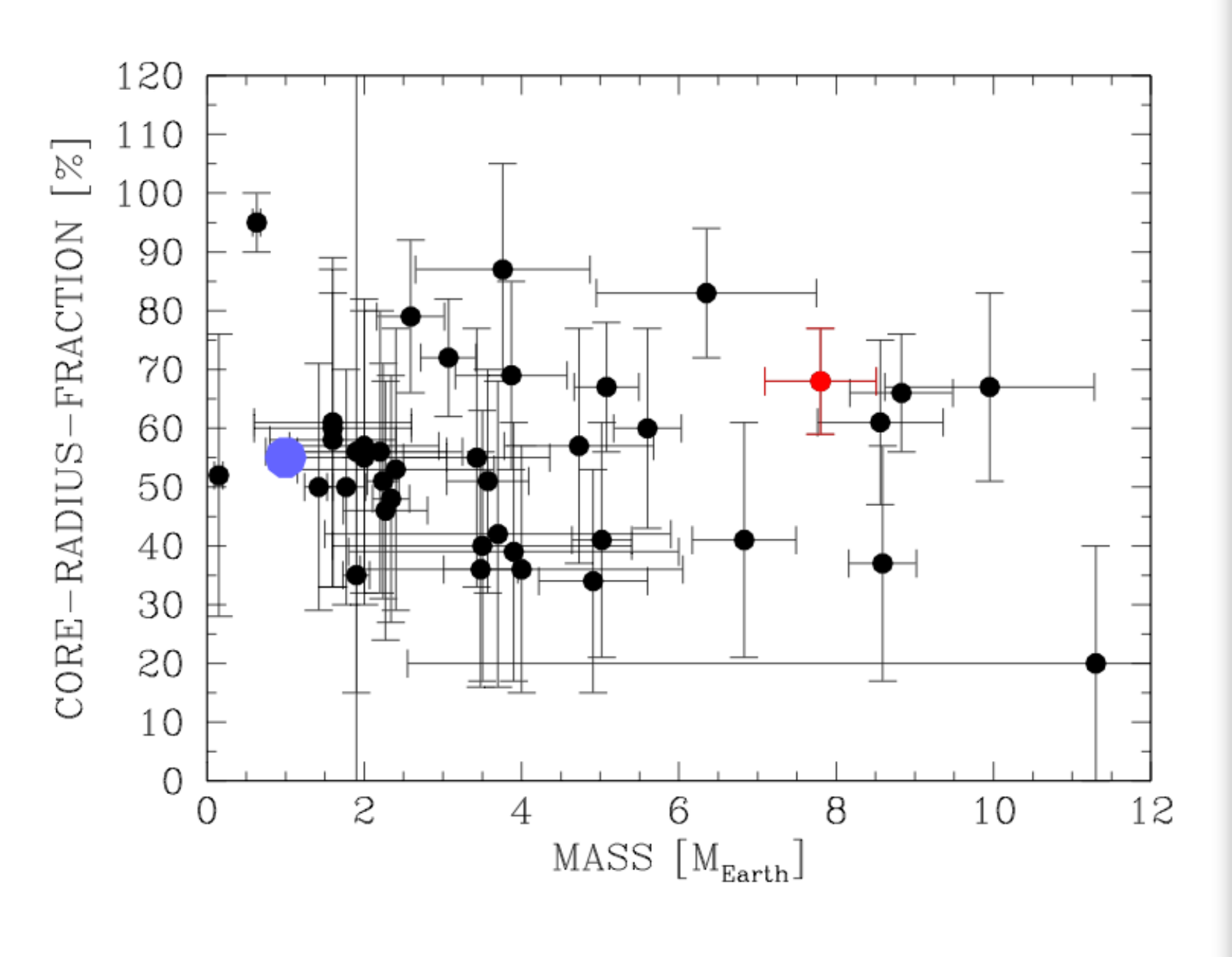}
\caption{The core-radius fraction versus the
mass of the planet. No correlation between the mass of the planet and
$\rm CRF$ is seen. The red point is K2-106\,b, the blue point the Earth.}
\label{Fig13}
\end{figure}

Using the HARDCORE model provided by NASA  \citep{2018MNRAS.476.2613S}, we also calculate the marginal core radii fraction (CRF). Figure \ref{Fig13} shows the marginal core radii fraction for all known lmUSPs. The red symbol is K2-106\,b. No relation between the marginal core radius fraction and the mass is seen. We also mark the position of the Earth as a blue dot, although the Earth is not an USP.

\begin{table*}
\caption{Properties of USPs with $Mp\leq 25\,M_{\oplus}$ that have mass and radius measurements}
\begin{tabular}{l l l l l l l}
\hline
\noalign{\smallskip}
name        & mass            & radius                  & period & a & CRF & references \\
            & [$M_{\oplus}$]  & [$R_{\oplus}$]          & [days] & [AU] & [\%] &           \\  \hline
TOI-731\,b    & $0.15_{-0.04}^{+0.07}$ & $0.59\pm0.02$  & 0.322 & 0.0069 & $52\pm 24$ & \citet{2022AJ....163...99G} \\           
KOI-4777$^{2}$      & $\leq 0.34$   & $0.51\pm0.03$           & 0.412  & 0.0069 & $<100$ & \citet{2022AJ....163....3C} \\ % 
GJ367\,b & $0.633\pm0.050$ & $0.699\pm0.024$ & 0.322 & 0.0071 & $95\pm5$ & \citet{2023ApJ...955L...3G} \\ % two planet system
GJ1252\,b     & $1.42\pm0.18$ & $ 1.166_{-0.058}^{+0.061}$ & 0.548 & 0.0128 & $ 50\pm21$ & \citet{2022NatAs...6..736S} \\ % three planet system
TOI-500\,b    & $1.6_{-0.7}^{+1.3}$ & $ 1.16\pm0.12$  & $0.548$ & 0.0128 & $ 61\pm28$ & \citet{2022AJ....163...99G} \\
TOI-1442\,b & $1.6_{-0.5}^{+1.1}$ & $1.17\pm0.06$  & 0.409 & 0.0071 & $58\pm25$ & \citet{2022AJ....163...99G} \\
TOI-2290\,b   & $1.6_{-0.6}^{+1.4}$ & $1.17\pm0.07$  & 0.386 & 0.0086 & $60\pm27$ & \citet{2022AJ....163...99G} \\
Kepler-78\,b  & $1.77_{-0.25}^{+0.24}$ & $1.228_{-0.018}^{+0.019}$ & 0.355 & 0.01 & $50\pm20$  & \citet{2019ApJ...883...79D} \\ % single
GJ806\,b & $1.90\pm0.17$  & $1.331\pm0.023$ & 0.926 & 0.0844 & $35\pm20$  & \citet{2023AuA...678A..80P} \\ % two planet system
TOI-539\,b   & $1.9_{-0.7}^{+1.6}$ & $1.25\pm0.10$  & 0.310 & 0.0089 & $56\pm26$ & \citet{2022AJ....163...99G} \\
TOI-833\,b &  $2.0_{-0.6}^{+1.5}$ & $1.27\pm0.07$ & $1.042^{(3)}$ & 0.0171 & $55\pm25$  & \citet{2022AJ....163...99G} \\ % single
TOI-2445\,b    & $2.0_{-0.7}^{+1.2}$ & $1.25\pm0.08$  & 0.371 & 0.0064 & $57\pm25$ & \citet{2022AJ....163...99G} \\
TOI-206\,b    & $2.2_{-0.7}^{+1.4}$ & $1.30\pm0.05$  & 0.736 & 0.0112 & $56\pm24$ & \citet{2022AJ....163...99G}  \\ % single
TOI-561\,b    & $2.24\pm0.20$ & $1.31\pm0.04$  & $ 1.066^(3)$ & 0.0204 & $51\pm20$ & \citet{2023AJ....165...88B} \\ % single
TOI-1807\,b   & $2.27_{-0.58}^{+0.49}$ & $1.37_{-0.09}^{+0.10}$  & 0.549 & 0.0135 & $46\pm22$ & \citet{2022arXiv221004162P} \\
LTT3780\,b &  $2.34_{-0.23}^{+0.24}$ & $1.35\pm0.06$ & 0.768 & 0.0120 & $48\pm21$ & \citet{2020AuA...642A.173N} \\
TOI-1263\,b    & $2.4_{-0.8}^{+1.7}$ & $1.35\pm0.06$  & $1.021^{(3)}$ & 0.0185 & $53\pm24$ & \citet{2022AJ....163...99G} \\
K2-229\,b     & $2.59\pm0.43$ & $1.165\pm0.066$ & 0.584 & 0.0131 & $79\pm13$ & \citet{2018NatAs...2..393S} \\ % (Mercury like)
TOI-431\,b    & $3.07\pm0.35$ & $1.28\pm0.04$ & 0.490     & 0.012 & $72\pm10$  & \citet{2021MNRAS.507.2782O} \\
TOI-1685\,b   & $3.43\pm0.93$   & $1.459\pm0.065$ & 0.669 & 0.0116 & $55\pm22$ & \citet{2021AJ....162..161H} \\
TOI-1416\,b  &  $3.48\pm0.47$ &  $1.62\pm0.08$ & $1.067^{(3)}$ & 0.0190 & $36\pm20$  & \citet{2023AuA...677A..12D}\\
TOI-2260\,b    & $3.5_{-1.3}^{+2.5}$ & $ 1.62\pm0.13$  & $0.352$ & 0.0097  & $40\pm23$ & \citet{2022AJ....163...99G} \\
Kepler-10\,b  & $3.57_{-0.53}^{+0.51}$ & $1.489_{-0.021}^{+0.023}$ & 0.837 & 0.0172 & $51\pm19$  & \citet{2019ApJ...883...79D} \\
TOI-1242\,b    & $3.7_{-1.5}^{+2.9}$ & $1.65\pm0.23$  & 0.381 & 0.0097 & $42\pm26$ & \citet{2022AJ....163...99G} \\
TOI-1238\,b   & $3.76_{-1.07}^{+1.15}$ & $1.21_{-0.10}^{+0.11}$ & 0.764 & 0.0139 & $87\pm18$  & \citet{2022AuA...658A.138G} \\
TOI-1444\,b   & $3.87\pm0.71$ & $1.397\pm0.064$ & 0.470 & - & $69\pm16$ & \citet{2021AJ....162...62D} \\
TOI-2411\,b$^{(4)}$ & $ 3.9_{-1.4}^{+2.8}$ & $1.68\pm0.11$ & 0.783 & 0.0144 & $ 39\pm22$ & \citet{2022AJ....163...99G} \\
TOI-1075\,b    & $4.0_{-1.4}^{+2.7}$ & $1.72\pm0.08$  & 0.604 & 0.0118 & $36\pm21$ & \citet{2022AJ....163...99G} \\
CoRoT-7\,b    & $4.73\pm0.95$   & $1.58\pm0.10$           & 0.854 & 0.0172 & $57\pm20$ & \citet{2014MNRAS.443.2517H} \\
TOI-1634\,b   & $4.91_{-0.70}^{+0.68}$ & $1.790_{-0.081}^{+0.080}$  & 0.989 & 0.0155 & $34\pm19$  & \citet{2021AJ....162...79C} \\	
HD3167\,b     & $5.02\pm0.38$   & $1.70_{-0.074}^{+0.18}$ & 0.960 & 0.0186 & $41\pm20$ & \citet{2017AJ....154..122C} \\
K2-141\,b     & $5.08\pm0.41$ & $1.51\pm0.05$ & 0.280 & - & $67\pm11$ & \citet{2018AJ....155..107M} \\
HD80653\,b    & $5.60\pm0.43$ & $1.613\pm0.071$ & 0.720 & 0.0166 & $60\pm17$  & \citet{2020AuA...633A.133F} \\ % Earth-like
Kepler-407\,b & $6.35\pm1.4$ & $1.43\pm0.03$ & 0.669 & - & $83\pm11$ & \citet{2014ApJS..210...20M} \\
WASP-47\,e    & $6.83\pm0.66$   & $1.810\pm0.027$ & 0.790 & 0.0173 & $41\pm20$  & \citet{2017AJ....154..237V} \\
K2-106\,b     & $7.80_{-0.70}^{+0.71}$ & $1.676_{-0.037}^{+0.037}$ & 0.571 & 0.0131 & $68\pm9$ & This article \\
55\,Cnc\,e    & $8.59\pm0.43$   & $1.947\pm0.038$         & 0.737 & 0.0154 & $37\pm20$ & \citet{2018RNAAS...2..172C} \\
HD213885\,b    & $8.83_{-0.65}^{+0.66}$ & $1.745_{-0.052}^{+0.051}$  & $1.008^{(3)}$ &  0.0201 & $66\pm10$ & \citet{2020MNRAS.491.2982E} \\
TOI-1075\,b   & $9.95_{-1.30}^{+1.36}$ & $1.791_{-0.116}^{+0.081}$ & 0.605 & 0.0118 & $67\pm16$ & \citet{2023AJ....165...47E} \\
K2-266\,b$^{(5)}$ & $11.3_{-6.5}^{+11}$ & $3.3_{-1.3}^{+1.8}$ & 0.658 & 0.0131 & $20_{-20}^{+30}$ &  \citet{2018AJ....156..245R} \\
% TOI-849\,b  &  $39.1_{-2.6}^{+2.7}$ & $ 3.45_{-0.12}^{+0.16}$ & $1.0697^{9}$ & 0.01598 & & \citet{2020Natur.583...39A}\\
\hline
\end{tabular}
\\
$^{(1)}$ Marginal core radius fraction (CRFmarg) \citep{2018MNRAS.476.2613S};
$^{(2)}$ Although the mass of this planet has not been determined yet, we include it in the table because the upper limit of the mass is very small; 
% $^{(3)}$ Several values are reported for this planet, because of the reported values are significantly different. 
$^{(3)}$ Strictly speaking this planet is not an USP. We also include planets
with orbital period between 1.0 and 1.1 days to make sure that we list all USPs even if there is still a small error of the period.
$^{(4)}$ TOI-2290\,b=TOI-2411\,b; 
$^{(5)}$ This planet has unusually large errors that puts it outside Fig.\ref{USPsMpMr}.\\
\label{tab:USPs}
\end{table*}

K2-106\,c has the same mass as K2-106\,b but a radius of $2.84_{-0.08}^{+0.10}$ $R_{\oplus}$.
This planet thus is likely to have an extended atmosphere.
Such planets are often called mini-Neptunes, which is a bit misleading.
They are not like Neptune, as their atmospheres contain only a few percent of the masses of the planets. For example, a Hydrogen rich atmosphere containing 1-2\% of the mass of K2-106\,c would fit the data \citep{2019PNAS..116.9723Z}. We thus prefer to call such objects C-class planets instead.

The K2-106-system is very interesting, because it contains two planets of almost the same mass but different density. What can we say about the possible formation scenarios?
Most of the UPSs in Table\,\ref{tab:USPs} are in multiple systems.  
Because hot Jupiters are lonely, it is unlikely that any of the planets in a system is a remnant core of a gas giant. Since most USPs are either in class-B, or class-C there is currently no
evidence that they have an unusual composition. Perhaps, they form just like planets
at larger distances.

The mass and radius of K2-106\,c is $\rm Mp=7.3_{-2.4}^{+2.5}$ $\rm M_{\oplus}$ and $\rm Rp=2.84_{-0.08}^{+0.10}$ $\rm R_{\oplus}$. The
radius thus is significantly larger than that of USPs with similar masses. This planet thus presumably has an hybrid or Hydrogen rich atmosphere.
The fact that one planet has an atmosphere, and the other does not, can be
explained with core-powered mass-loss \citep{2013ApJ...776....2L,2018MNRAS.476..759G}, or
atmospheric evaporation due to the XUV-radiation from the host star
\citep{2007A&A...472..329E,2017AuA...598A..90F, 2018A&A...619A.151K,2018MNRAS.477..808L,2018A&A...619A.151K,2021MNRAS.500.4560P}.
More complicated mechanisms are not needed to explain why the inner planet does not have an extended atmosphere whereas the outer planet does, even if both planets formed from similar material.

\section*{Acknowledgements}

Based on observations collected at the European Southern Observatory under ESO programme 0103.C-0289(A). We are very thankful to the ESO-staff for carrying out the observations in service mode, and for providing the community with all the necessary tools for reducing and analysing the data.  

This paper includes data collected by the Kepler mission and obtained from the MAST data archive at the Space Telescope Science Institute (STScI). Funding for the Kepler mission is provided by the NASA Science Mission Directorate. STScI is operated by the Association of Universities for Research in Astronomy, Inc., under NASA contract NAS 5–26555.

This paper includes data collected by the TESS mission.
Funding for the TESS mission is provided by NASA’s Science Mission Directorate. We acknowledge the use of public TOI Release data from pipelines at the TESS Science Office and at the TESS Science Processing Operations Center. 

This work has made use of data from the European Space Agency (ESA) mission Gaia (https://www.cosmos. esa.int/gaia), processed by the Gaia Data Processing and Analysis Consortium (DPAC, https://www.cosmos.esa. int/web/gaia/dpac/consortium). Funding for the DPAC has been provided by national institutions, in particular the institutions participating in the Gaia Multilateral Agreement. The Gaia mission website is https://www.cosmos. esa.int/gaia https://www.cosmos.esa.int/gaia. The Gaia archive website is https://archives.esac.esa.int/gaia. 

This publication makes use of data products from the Two Micron All Sky Survey, which is a joint project of the University of Massachusetts and the Infrared Processing and Analysis Center/California Institute of Technology, funded by the National Aeronautics and Space Administration and the National Science Foundation

This work has made use of the Mikulski Archive for Space Telescopes (MAST). 
MAST is a NASA funded project to support and provide to the astronomical community a variety of astronomical data archives, with the primary focus on scientifically related data sets in the optical, ultraviolet, and near-infrared parts of the spectrum.

This research has made use of the SIMBAD database, operated at CDS, Strasbourg, France.

This work was generously supported by the Deutsche Forschungsgemeinschaft (DFG) in the framework of the priority programme ``Exploring the Diversity of Extrasolar Planets'' (SPP 1992) in program GU 464/22, and by the Th\"uringer Ministerium f\"ur Wirtschaft, Wissenschaft und Digitale Gesellschaft.  

E.~G. acknowledges the generous support from the Deutsche Forschungsgemeinschaft (DFG) of the grant HA3279/14-1.

J.~K. gratefully acknowledges the support of the Swedish National Space Agency (SNSA; DNR 2020-00104) and of the Swedish Research Council  (VR; Etableringsbidrag 2017-04945).

This research is also supported work funded from the European Research Council (ERC) the European Union’s Horizon 2020 research and innovation programme (grant agreement n◦803193/BEBOP).

\section*{Data Availability Statement}

The data underlying this article are available in the ESO Science
Archive Facility http://archive.eso.org/cms.html and from the 
Barbara A. Mikulski Archive for Space Telescopes (MAST).

%%%%%%%%%%%%%%%%%%%%%%%%%%%%%%%%%%%%%%%%%%%%%%%%%%



%\bibliographystyle{mnras}


% \bibitem[\protect\citeauthoryear{Author}{2012}]{Author2012}
% Author A.~N., 2013, Journal of Improbable Astronomy, 1, 1

% \bibitem[\protect\citeauthoryear{Others}{2013}]{Others2013}
% Others S., 2012, Journal of Interesting Stuff, 17, 198

% xxxxxxxxxxxxxxxxxxxxxxxxxxxxxxxxx

\bibliographystyle{mnras}
\bibliography{literature} % if your bibtex file is called example.bib

%%%%%%%%%%%%%%%%%%%%%%%%%%%%%%%%%%%%%%%%%%%%%%%%%%

%%%%%%%%%%%%%%%%% APPENDICES %%%%%%%%%%%%%%%%%%%%%

\appendix
\section{Posterior parameters}

%\begin{figure*}
%\includegraphics[height=1.35\textwidth,angle=0.0]{Images/Fig10.pdf}
%\caption{Posterior parameters.}
%\label{post}
%\end{figure*}

%\begin{figure*}%[ht!]
%     \centering
%     \begin{subfigure}[b]{0.9\textwidth}
%         \centering
%         \includegraphics[width=\textwidth]{Images/K2-106_posteriorA.png}
%         \label{}
%     \end{subfigure}
 %    \begin{subfigure}[b]{0.9\textwidth}
%         \centering
%         \includegraphics[width=\textwidth]{Images/K2-106_posteriorB.png}
%         \label{}
%     \end{subfigure}
%        \caption{Posterior distributions of the fitted parameters for the best-fit model discussed in section\,\ref{sectIVa}.}
%        \label{fig:posteriors}
%\end{figure*}

\begin{figure*}
\centering
    \includegraphics[width=0.9\textwidth, angle=0]{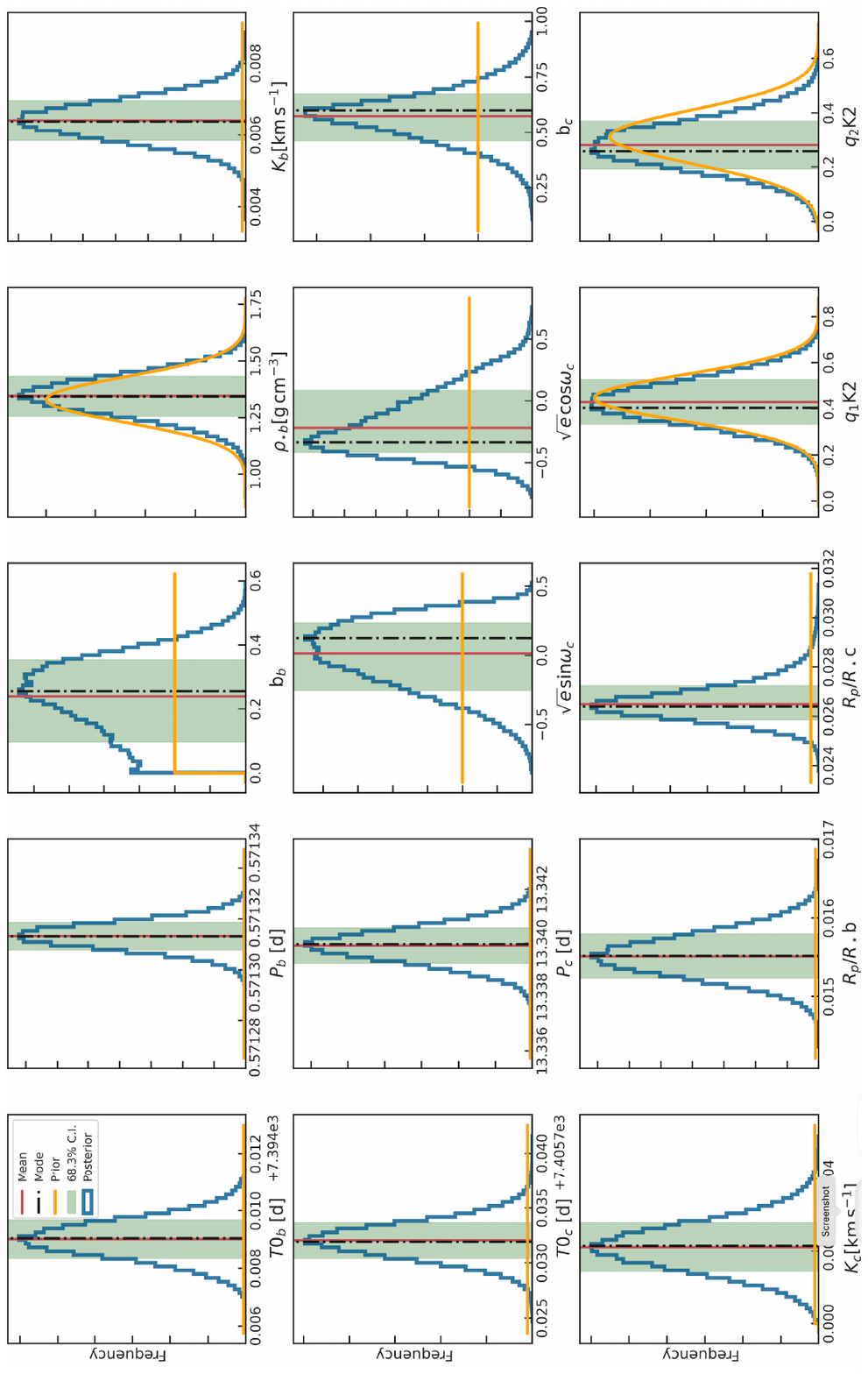}
    \caption{Posterior distributions of the fitted parameters for the best-fit model discussed in section\,\ref{sectIVa} (Continued on next page).}
    \label{fig:posteriors}
\end{figure*}

\begin{figure*}
\centering
    \includegraphics[width=0.9\textwidth, angle=0]{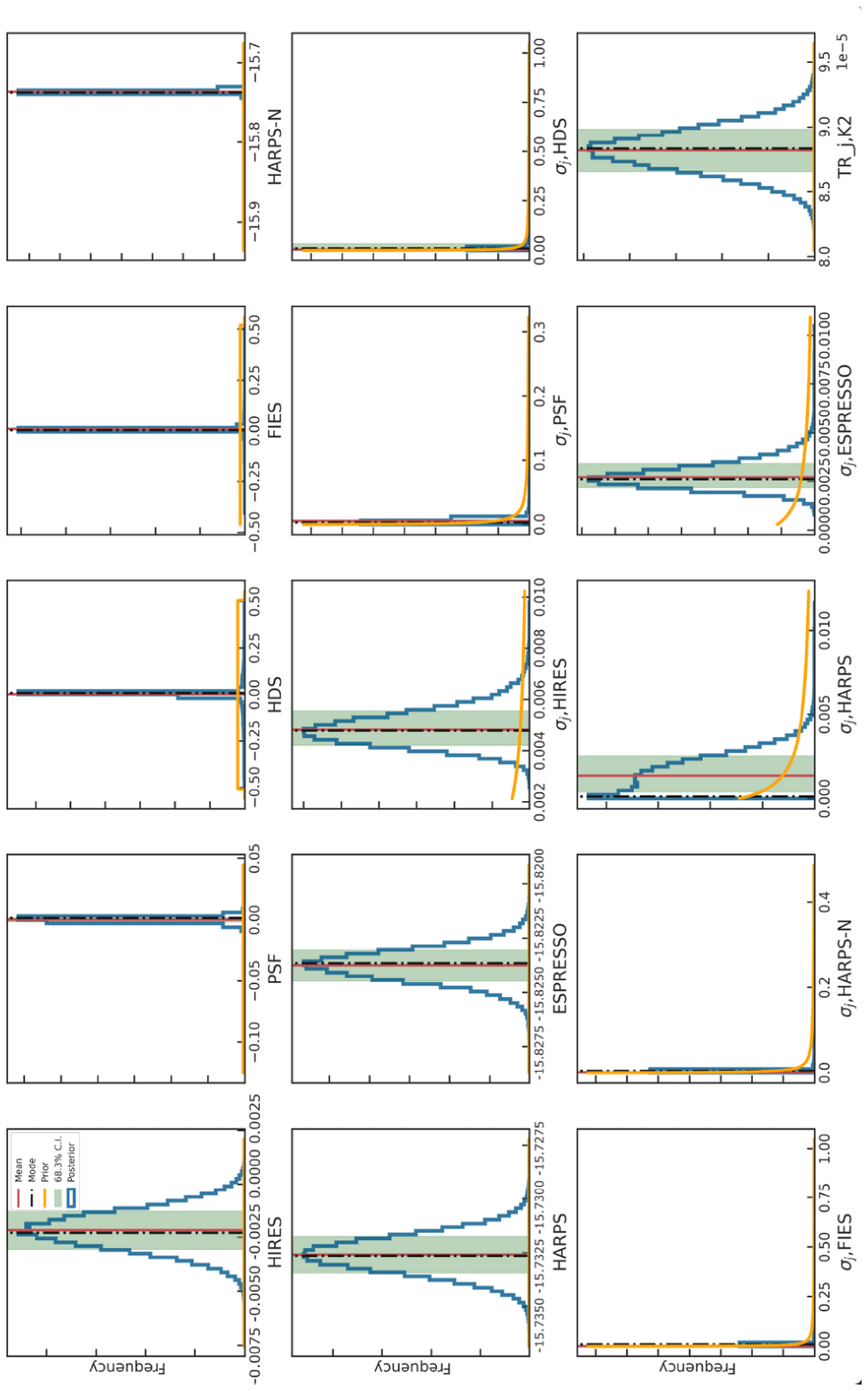}
    \caption{Posterior distributions of the fitted parameters for the best-fit model discussed in section\,\ref{sectIVa}.}
    \label{fig:posteriorsB}
\end{figure*}


%%%%%%%%%%%%%%%%%%%%%%%%%%%%%%%%%%%%%%%%%%%%%%%%%%

% Don't change these lines
\bsp	% typesetting comment
\label{lastpage}
\end{document}